\documentclass[12pt,noshowpacs,nofootinbib,notitlepage,amsmath]{revtex4-1}
\usepackage{setspace}
\allowdisplaybreaks
\usepackage{graphicx,color}
\usepackage[colorlinks=true,citecolor=blue,linkcolor=blue,urlcolor=blue]{hyperref}
\usepackage[charter]{mathdesign}
\usepackage{multirow}
\usepackage{enumerate}

\DeclareSymbolFontAlphabet{\mathcal}{symbols}
\DeclareSymbolFont{symbols}{OMS}{xmdcmsy}{m}{n}
\DeclareSymbolFont{largesymbols}{OMX}{xmdcmex}{m}{n}
\SetSymbolFont{symbols}{bold}{OMS}{xmdcmsy}{b}{n}

\begin{document}  
\title{\color{blue}\Large Gravitational Wave ``Echo'' Spectra}

\author{Randy S. Conklin}
\email{rconklin@physics.utoronto.ca}
\author{Bob Holdom}
\email{bob.holdom@utoronto.ca}
\affiliation{Department of Physics, University of Toronto, Toronto, Ontario, Canada  M5S 1A7}
\begin{abstract}
Exotic compact objects may resemble black holes very closely while remaining horizonless. They may be distinguished from black holes because they effectively give rise to a resonant cavity for the propagation of low frequency gravity waves. In a Green's function approach, the resonance structure appears in a transfer function. The transfer function in turn is modulated by an initial-condition-dependent source integral to obtain the observed spectrum. We find that the source integral displays universal factors that tend to enhance low and negative frequencies, and this increases the complexity of the waveforms in the time domain. These waveforms also display a significant sensitivity to initial conditions. For these reasons a standard matched-filter search strategy is difficult. In contrast, the sharp and evenly spaced resonance spectrum presents a much more robust signal to target. It persists even in the absence of simple echoes. We also describe an additional two-component structure of this resonance pattern.\end{abstract}
\maketitle

\section{Introduction}\label{Introduction}
The Einstein action has made us very accustomed to black holes and their ``no drama'' event horizons. But the Einstein action should eventually be subsumed into a UV complete theory of gravity, and in such a theory there may well be new horizonless solutions that in many respects appear much like black holes. Within a Planck length of the would-be horizon, strong gravity and high curvatures quickly turn on. These objects are very close to being completely black, but not quite. An old candidate for a UV complete theory of gravity is quadratic gravity, long known to be renormalizable and asymptotically free. The not quite black hole solutions found in this theory are called 2-2-holes \cite{Holdom:2016nek}.

An ideal probe to test for not quite black holes are the low frequency gravitational waves that are excited in and around them when they are newly formed, as in the merger events observed by LIGO. Waves can travel in and out of a 2-2-hole in a finite time if they are not scattered or absorbed by anything inside the 2-2-hole, which is likely to be the case only for gravitational waves. These waves also do not encounter an angular momentum barrier at the origin, which is present for more normal compact objects. For this reason a good approximation to a wave equation around a 2-2-hole is a wave equation around a truncated Schwarzschild background, which has a reflecting wall at a radius just lightly outside the horizon. A truncated Kerr black hole can be considered similarly. On such a background one can study the standard gravitational wave equation, and in particular the radial equation. In the Schwarzschild case this is characterized by a potential~\cite{Berti:2009kk}, which as usual is dominated by an angular momentum barrier that peaks near the light ring radius of $r=3M$. With the wall also in place, the result then is a 1D cavity bounded by the wall and the potential barrier, such that low frequency perturbations are nearly confined within the cavity. The wall represents the origin ($r=0$) of a not quite black hole. Since the barrier is finite the perturbation will slowly leak out, and in certain circumstances this can lead to pulses being emitted at regular intervals~\cite{Cardoso:2016rao,Cardoso:2016oxy}. 

In this simplest case a distant observer may then detect ``echoes,'' and it is these well defined echoes in the time domain that have been the focus as a search target. Every subsequent echo tends to broaden and eventually the individual echoes merge and are no longer distinguishable. Even the presence of the early echoes requires a simple enough initial perturbation that can be thought to bounce back and forth within the cavity. But there is a more general property of the waveform that is evident in its frequency content. Due to the nearly trapped modes of a cavity, the spectrum features sharp resonances. These quasinormal modes of the system differ markedly from the QNMs of a normal black hole. The simple 1D cavity structure of the radial equation means that the spacing between resonances is very nearly constant throughout the spectrum and is given by $\Delta f=1/(2\Delta x)$, where $\Delta x$ is the size of the cavity in tortoise coordinates. When there are echoes then the time delay between echoes is $\Delta t=1/\Delta f$. A direct resonance search, rather than an echo search, was proposed and carried out in~\cite{Conklin:2017lwb}. The search involves taking the absolute value of the Fourier transform of the strain data as measured by LIGO. One of our goals in this paper is to explore the sensitivity of the resonance spectrum to initial conditions.

We are considering gravitational waves on a Kerr background and we shall focus on the dominant $\ell=2$, $m=2$ mode. The wave reaching us can be taken to be a complex amplitude to describe the two polarizations, and in our case of interest we can write
\begin{align}
h_{22}(t,\Omega)=h_+(t,\Omega)-ih_\times(t,\Omega)=\psi(t)Y_{22}(\Omega).
\end{align}
It is the frequency space counterpart of $\psi(t)$, which we will label $\psi_\omega$, that will be obtained from the appropriate wave equation. Since $\psi(t)$ is complex, $\psi_\omega$ is not conjugate symmetric about $\omega=0$. That is, the behaviour of $\psi_\omega$ for both positive and negative frequencies is needed to describe the two polarizations. The consideration of both positive and negative frequencies plays an important role in the following. The wave equation will have a source term $S(\omega,x)$ and will involve the tortoise coordinate $x$. Then $\psi_\omega$ can be extracted from a solution $\psi(\omega,x)$ by taking the limit $\psi(\omega,x)\to \psi_\omega e^{i\omega x} $ for $x\to\infty$.

The Green's function method developed in \cite{Conklin:2017lwb} leads to the following result,
\begin{align}
\psi_\omega &= K(\omega)D(\omega),\nonumber\\
K(\omega)&=\frac{1}{W(\omega,\infty)},\nonumber\\
D(\omega)&=\int_{-\infty}^\infty p(\omega,x')\psi_\textrm{left}(\omega,x') S(\omega,x') dx',\nonumber\\
p(\omega,x) &= \frac{W(\omega,\infty)}{W(\omega,x)}.
\label{eq:obspsi}
\end{align}
$K(\omega)$ is the transfer function and $D(\omega)$ is the source integral. $W(\psi_{\rm left},\psi_{\rm right})$ is the Wronskian where $\psi_\textrm{left}$ and $\psi_\textrm{right}$ are the solutions to the homogeneous wave equation. These solutions satisfy a boundary condition on the left and right respectively, with the former condition related to the reflectivity of the wall and the latter to the outgoing plane wave condition at large $x$. Note that due to the form of the wave equation (see (\ref{eq:SN}) below), it is $pW$ rather than $W$ that is independent of $x$.

$K(\omega)$ exhibits the resonance structure that is of interest as a search target. An alternative expression \cite{Conklin:2017lwb} for it involves the transmission and reflection amplitudes for a Kerr black hole,
\begin{align}
K(\omega)=\frac{T_{\rm Kerr}(\omega)}{1+R_{\rm Kerr}(\omega)R_{\rm wall}e^{2i(\omega-\omega_0)\Delta x}}.
\end{align}
A Schwarzschild version of this is well known \cite{Mark:2017dnq}. This form makes explicit the dependence on both the cavity size $\Delta x$ and $\omega_0=m\chi/2r_+$, with $\chi$ the dimensionless spin parameter, $r_+$ the horizon radius and $m=2$ in our case. $R_{\rm wall}$ specifies the boundary condition as discussed below. Our goal here is to understand more completely how $\psi_\omega$ is influenced by $D(\omega)$. We calculate $D(\omega)$ explicitly from initial conditions, and thus determine how the transfer function is modulated to produce the observable spectrum. At the same time we will see that $D(\omega)$ can introduce a significant variability into the time-domain waveform. 

\section{Analytical Modulation}\label{AMod}

The gravitational perturbations of interest are described by the Teukolsky equation for $s=-2$, and in particular the Teukolsky radial equation \cite{Teuk}. But this equation presents numerical difficulties since it possesses a long-range potential. Sasaki and Nakamura (SN) developed an equivalent equation possessing a short-range potential, and which also reduces to the standard Regge-Wheeler equation in the spinless limit. The SN equation takes the form \cite{Sasaki:1981sx}
\begin{equation}\label{eq:SN}
\frac{d^2X}{dx^2} - {\cal F}(\omega,x)\frac{dX}{dx} - {\cal U}(\omega,x)X = S(\omega,x).
\end{equation}
The $\cal F$ and $\cal U$ functions are nontrivial, but they are short-ranged, meaning that far enough away from $x=0$ they vanish or tend to constants.

The source term $S(\omega,x)$ shall embody the initial conditions, and we obtain it by applying the Laplace transform to the time-dependent wave equation. The Teukolsky radial wave equation can be written as such, but the $\omega$ dependence of the SN equation in (\ref{eq:SN}) is sufficiently complex that the connection to a time-dependent wave equation is lost. Our approach for obtaining $S(\omega,x)$ will be to consider the SN equation in the asymptotic limits away from the potential. In these limits the corresponding wave equations are obvious and thus $S(\omega,x)$ can be determined via Laplace transform. In fact we shall develop initial conditions corresponding to travelling pulse solutions of the wave equations that then travel to the potential barrier in a known way. This allows us to construct and explore the effect of the source integral $D(\omega)$ in sufficient generality.\footnote{Pulses that travel towards the potential barrier, either from the inside or outside, also serve as initial conditions for the normal black hole. In this case a single pulse involving the black hole QNMs is emitted, thus modelling the ringdown phase of a merger event. For a not quite black hole, the initial pulse is the same as for a black hole \cite{Cardoso:2016rao,Cardoso:2016oxy}, and so the modelling of the ringdown is the same. Only the subsequent echoes are sensitive to the presence of the wall.}  

Sufficiently far from the potential the homogeneous SN equation reduces to
\begin{align}
\frac{d^2X}{dx^2} - \omega^2 X = 0\quad\textrm{and}\quad\frac{d^2X}{dx^2} - (\omega-\omega_0)^2 X = 0,\label{e3}
\end{align}
on the outside and inside of the cavity, respectively. The corresponding wave equations are
\begin{align}
\partial^2_x\psi - \partial^2_t\psi= 0\quad\textrm{and}\quad
\partial^2_x\psi - \partial^2_t\psi + 2i\omega_0\partial_t\psi + \omega_0^2\psi = 0.
\label{e1}\end{align}
The plane wave solutions are $e^{-i\omega (t \pm x)}$ and $e^{-i\omega t \pm i(\omega-\omega_0) x}$ respectively, and we shall build up travelling pulses by taking superpositions of these plane waves.

Our main focus is on initial conditions corresponding to pulses starting inside the cavity, but we shall also consider outside pulses below. For the inside pulses we consider both outgoing and ingoing ones as follows,
\begin{align}
\psi_\rightarrow(t,x) = \int_{-\infty}^\infty |b_0(\omega)| f_\rightarrow(\omega) e^{-i\omega(t-t_0)+i (\omega-\omega_0) (x-x_1)} d\omega,\nonumber\\
\psi_\leftarrow(t,x) = \int_{-\infty}^\infty |C(\omega)| f_\leftarrow(\omega) e^{-i\omega(t-t_0)-i (\omega-\omega_0) (x-x_2)} d\omega.
\label{e6}\end{align}
$x_1$ and $x_2$ are the initial positions of the two pulses at time $t_0$. The functions $|C(\omega)|$ and $b_0(\omega)$ (and $c_0(\omega)$ appearing below) are defined in the Appendix, where $b_0$ and $c_0$ are related to the transformation of Teukolsky to SN amplitudes.  With their presence, the results in the Appendix show that the $f_{\leftarrow\atop \rightarrow}$'s are directly related to the spectral flux density of the pulse,
\begin{align}
\frac{dE_{\leftarrow\atop \rightarrow}}{d\omega}=8\omega(\omega-\omega_0)|f_{\leftarrow\atop \rightarrow}(\omega)|^2.
\label{e4}\end{align}
Implicit in (\ref{e6}) is that the integrals converge. We typically take $f_{\leftarrow\atop \rightarrow}(\omega)$ to be Gaussian, in which case $\psi_{\leftarrow\atop \rightarrow}$ are localized pulses that travel with constant absolute value and speed (in the tortoise coordinate and with the ingoing pulse reflecting off the wall) until they enter the potential barrier region where the evolution is governed by the full SN equation. From (\ref{e4}) we see that the flux densities are negative in the superradiance region $0<\omega<\omega_0$. 

The source term $S(\omega,x)$ can then be obtained from initial conditions as specified by these pulses at time $t_0$. The Laplace transform is defined by
\begin{equation}
\psi(\omega,x)=\int_{t_0}^{\infty}\psi(t,x) e^{i\omega t}dt,
\end{equation}
which when applied to (\ref{e1}) gives $t_0$-dependent terms that define $S(\omega,x)$,
\begin{equation}
S(\omega,x) = i(\omega - 2\omega_0)\psi(t_0,x)- \partial_t\psi(t,x)|_{t=t_0}.
\label{e2}\end{equation}
Then inserting our pulses we have $S(\omega,x)=S_\leftarrow(\omega,x)+S_\rightarrow(\omega,x)$ with
\begin{align}
S_\rightarrow(\omega,x)=\int_{-\infty}^\infty (\omega+\omega^\prime - 2\omega_0)|b_0(\omega)| f_\rightarrow(\omega^\prime) e^{i (\omega^\prime -\omega_0)(x-x_1)} d\omega^\prime,\nonumber\\
S_\leftarrow(\omega,x)=\int_{-\infty}^\infty (\omega+\omega^\prime - 2\omega_0)|C(\omega)| f_\leftarrow(\omega^\prime) e^{-i (\omega^\prime -\omega_0)(x-x_2)} d\omega^\prime.
\end{align}

We can now proceed to evaluate the source integral $D(\omega)$ defined in (\ref{eq:obspsi}),
\begin{equation}
D(\omega)=\int_{-\infty}^{\infty} p(\omega,x)\psi_\textrm{left}(\omega,x) S(\omega,x) dx.
\end{equation}
We are considering initial positions of the pulses that are well away from the potential, so that $S(\omega,x)$ only has support well away from the potential. For such $x$, $p(\omega,x)$ becomes constant $p(\omega,x)\rightarrow p(\omega)$. Each term in
\begin{align}
\psi_\textrm{left}(\omega,x)  = A_\textrm{trans}(\omega)e^{-i (\omega-\omega_0) x}+A_\textrm{ref}(\omega)e^{i (\omega-\omega_0) x}
\end{align}
then produces delta functions via the $x$ integral, $\delta(\omega + \omega^\prime - 2\omega_0)$ and $\delta(\omega - \omega^\prime)$. We see that only the latter gives a nonvanishing contribution, and that it produces a factor of $2(\omega-\omega_0)$.
The result is then
\begin{equation}
D(\omega)=2i p(\omega)(\omega - \omega_0)(|b_0(\omega)|f_\rightarrow(\omega)A_\textrm{trans}e^{-i (\omega-\omega_0) x_1}+|C(\omega)|f_\leftarrow(\omega)A_\textrm{ref}e^{i (\omega-\omega_0) x_2}).
\end{equation}

The wall at location $x_w$ imposes the boundary condition satisfied by $\psi_\textrm{left}$. (It is convenient to associate $x=0$ with $r=3M$, in which case the location of the wall is $x_w=-\Delta x$.) This boundary condition is specified by $R_{\rm wall}(\omega)$, such that $|R_{\rm wall}|^2$ is the ratio of the outgoing and ingoing energy fluxes near the wall. $R_{\rm wall}=-1$ is the analog of the Dirichlet boundary condition while $R_{\rm wall}=0$ is the standard black hole boundary condition. To agree with the standard normalization in the latter case we then have
\begin{equation}
\psi_\textrm{left}(\omega,x)=\frac{e^{-i(\omega-\omega_0)x_w}}{2i\sqrt{\omega(\omega-\omega_0)}}(e^{-i(\omega-\omega_0)(x - x_w)} + R_{\rm wall}(\omega)R(\omega) e^{i(\omega-\omega_0)(x - x_w)}).
\end{equation}
$R(\omega)$ is a known real and nonnegative spin-dependent function with $R(\omega_0)=1$. The normalization of $\psi_\textrm{left}$ affects both the transfer function and the source integral in a way that cancels out. In this sense the factors of $\sqrt{\omega(\omega-\omega_0)}$ are superfluous, and we shall redefine the transfer function and the source integral in the following such that $K(\omega)\rightarrow K(\omega)/\sqrt{\omega(\omega-\omega_0)}$ and $D(\omega) \rightarrow D(\omega)\sqrt{\omega(\omega-\omega_0)}$. These redefined quantities display simpler behavior for $\omega$ near 0 and $\omega_0$.

We can now use the following results that were obtained in \cite{Conklin:2017lwb},
\begin{align}
p(\omega) &= \frac{|c_0(\omega)|}{|b_0(\omega)|}e^{i\phi(\omega)},\\
R(\omega) &= \frac{|b_0(\omega)|}{|C(\omega)|}.
\label{e8} \end{align}
$\phi(\omega)$ is a slowly varying phase. Factors of $b_0$ and $C$ cancel and we are left with
\begin{equation}\label{e5}
D(\omega)=(\omega - \omega_0)|c_0(\omega)|e^{i\phi}(f_\rightarrow(\omega)e^{-i (\omega-\omega_0) x_1}+R_{\rm wall} f_\leftarrow(\omega)e^{i (\omega-\omega_0) (x_2 - 2x_w)}).
\end{equation}
This can easily be generalized to a sum of ingoing and outgoing pulses at various initial positions $x_i$ inside the cavity.

We now consider an initial perturbation that originates on the outside of the cavity, that is outside the light ring. Now we need only consider ingoing pulses, and so 
\begin{align}
\psi(t,x) = \int_{-\infty}^\infty |C(\omega)| g(\omega) e^{-i\omega t-i \omega (x-x_3)} d\omega.
\end{align}
With this definition, the relation to the spectral flux density of the pulse is
\begin{align}
\frac{dE}{d\omega}=8\omega^2|g(\omega)|^2.
\label{e10}\end{align}
$S(\omega,x)$ is as in (\ref{e2}) with $\omega_0=0$ and $\psi_\textrm{left}$ on the outside of the potential takes the form
\begin{align}
\psi_\textrm{left}(\omega,x) = A_\textrm{in}(\omega)e^{-i \omega x}+A_\textrm{out}(\omega)e^{i \omega x}.
\end{align}
We thus end up with the result
\begin{align}
D(\omega)=2i\omega A_\textrm{out}(\omega)|C(\omega)|g(\omega)e^{i \omega x_3}.
\end{align}

Here we have what appears to be a surprising result. In the case of a perfectly reflecting wall (eg.~$R_{\rm wall}=-1$), it turns out that $|A_\textrm{out}(\omega)/A_\textrm{in}(\omega)|=|c_0(\omega)/C(\omega)|$. As seen in \cite{Conklin:2017lwb}, this is simply the statement that the ingoing and outgoing components of $\psi_\textrm{left}$ have identical flux densities in this special case. Since the transfer function can be written as $K(\omega)= 1/(2i\omega A_\textrm{in}(\omega))$, we find that $|\psi_\omega|= |K(\omega)D(\omega)|= |c_0(\omega)|g(\omega)$. This is a smooth slowly varying function for real $\omega$ and so the resonance structure has been completely eliminated. On the complex plane, for every pole in $K(\omega)$ there is now a zero from $D(\omega)$ in the complex conjugate position. We shall return to the consequences of this result (for an outside perturbation and perfect reflection) at the end of the next section.

\section{Consequences}

In this section we shall present examples where we give the signal $\psi_\omega$ and the corresponding waveform in the time domain. For the latter we employ the inverse of the Laplace transform,
\begin{equation}
\psi(t) = \frac{1}{2\pi}\int_{-\infty + i\varepsilon}^{\infty + i\varepsilon}  \psi_\omega\; e^{-i\omega t} d\omega.
\label{e11}\end{equation}
$\psi_\omega$ is viewed as a function on the complex plane and the value of $\varepsilon$ is chosen such that the integration contour is above all poles of $\psi_\omega$.  This contour is implicitly closed off in the lower half-plane and the result is independent of the value of $\varepsilon$. Practically we perform an integral over a finite range of $\omega$ and so $\varepsilon$ must be kept small. Poles that happen to lie in the upper half-plane correspond to growing modes in the time domain, and the inverse Laplace transform (unlike the Fourier transform) properly captures this growing-in-time behaviour. 

Shifting the integration variable by $\omega \rightarrow \omega + i\varepsilon$ gives
\begin{equation}
\psi(t) = \frac{e^{\varepsilon t}}{2\pi}\int_{-\infty}^{\infty}\psi_{\omega+i\varepsilon}e^{-i\omega t} d\omega .
\end{equation}
Thus computationally we perform an inverse Fast Fourier Transform (FFT) on a shifted frequency function. The sensitivity to this shift is almost entirely due to the shift in the transfer function $K(\omega)\rightarrow K(\omega+i\varepsilon)$, since the resonance heights are determined by how far the poles are offset from the contour. This shifted transfer function differs from transfer functions displayed in our earlier work \cite{Conklin:2017lwb}, in addition to the normalization change already mentioned. The narrowness of the resonance spikes is now limited, and thus the extreme sensitivity of $K(\omega)$ to the resolution (the frequency step size $\Delta\omega$) we observed before is eliminated. We choose $\varepsilon=\Delta\omega=2\times10^{-5}$.

The growing in time behavior is the ergoregion instability, which can be characterized by a slight amplification factor $1+Z(\omega)$ of each subsequent echo, with $Z(\omega)\lesssim .001$ for spin $\chi=2/3$ \cite{Brito:2015oca,Nakano:2017fvh, Conklin:2017lwb}. In turn this growing behaviour can be eliminated by a small amount of dissipation of the gravitational wave as it traverses the interior of the not quite black hole. This damping due to dissipation has been studied in \cite{Maggio:2017ivp,Maggio:2018ivz}. Here we model the required dissipation by adjusting the boundary condition to $R_{\rm wall}=-1+\epsilon$.\footnote{$R_{\rm wall}=-1$ is the boundary condition at the origin of a 2-2-hole and the dissipation may be induced by the relativistic gas that sources the 2-2-hole.} In the following we take $R_{\rm wall}=-0.995$, a value such that one or two hundred echoes are still contributing to a sharp resonance pattern. Of course as $|R_{\rm wall}|$ is reduced further, the echoes more quickly die away and the resonances broaden and shrink. Here we are interested in exploring the more striking signal.

\begin{figure}[h]
\centering
 \includegraphics[width=0.8\textwidth]{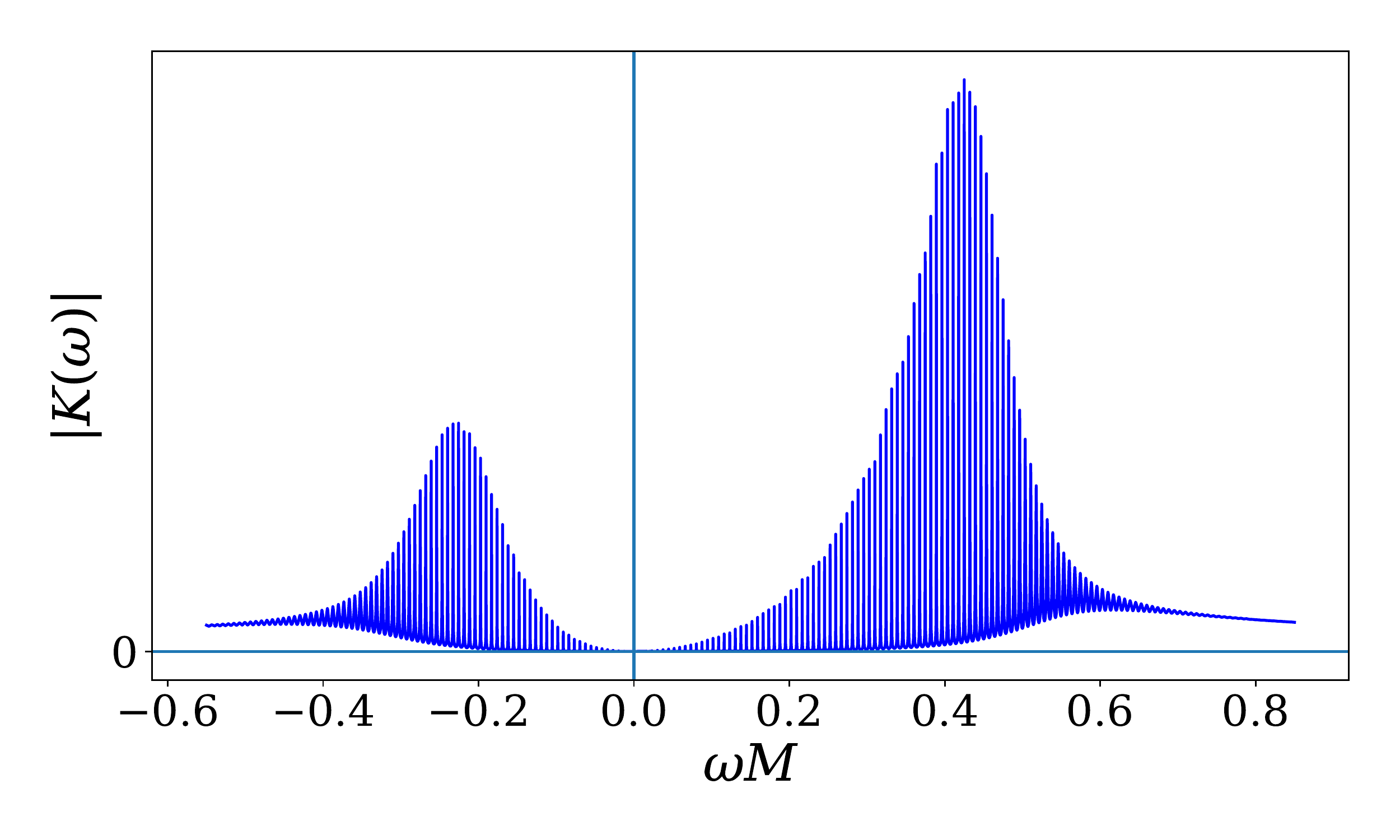}  
\caption{Absolute value of the transfer function for spin $\chi=2/3$, $l=m=2$ and $\Delta x=400$.}
\label{fig2}
\end{figure}

We first present the transfer function $|K(\omega+i\varepsilon)|$ in Fig.~\ref{fig2}. The spin $\chi=2/3$ implies $\omega_0=0.3820$, and for the cavity size we choose $\Delta x=400$. Both the positive and negative frequency parts of the spectrum are shown, with both parts needed to describe the two polarizations, as mentioned above.

Once $\psi_\omega=D(\omega)K(\omega)$ is obtained, the signal spectrum is exhibited in $|\psi_\omega|$. Our examples will also display the ``reconstructed'' spectrum, which is closer to an experimentally determined quantity. We have already mentioned how the complex waveform in the time domain describes the two polarizations. But the strain data from each detector is a set of real numbers. A projection from complex to real has been made that depends on the orientation of the detector. To model the action of a detector we shall take the real part of our derived signal waveform in the time domain. Taking the imaginary part or some other projection will give different but qualitatively similar results. We also truncate the time waveform to include some number of echoes. We then take the FFT of the real and truncated time series, and this is what we refer to as the reconstructed $\psi_\omega$.

Since the reconstructed $\psi_\omega$ is a FFT of a real series, the result is conjugate symmetric about $\omega=0$, meaning that the negative frequency part is redundant and so we need only consider positive frequencies. The result is that the reconstructed spectrum $|\psi_\omega|$ (at positive frequencies) has two components that come from the positive and negative frequency parts of the original spectrum. Our result for $D(\omega)$ will in turn help to determine the relative sizes of these two componennts. 

\begin{figure}[h]
\centering
 \includegraphics[width=0.6\textwidth]{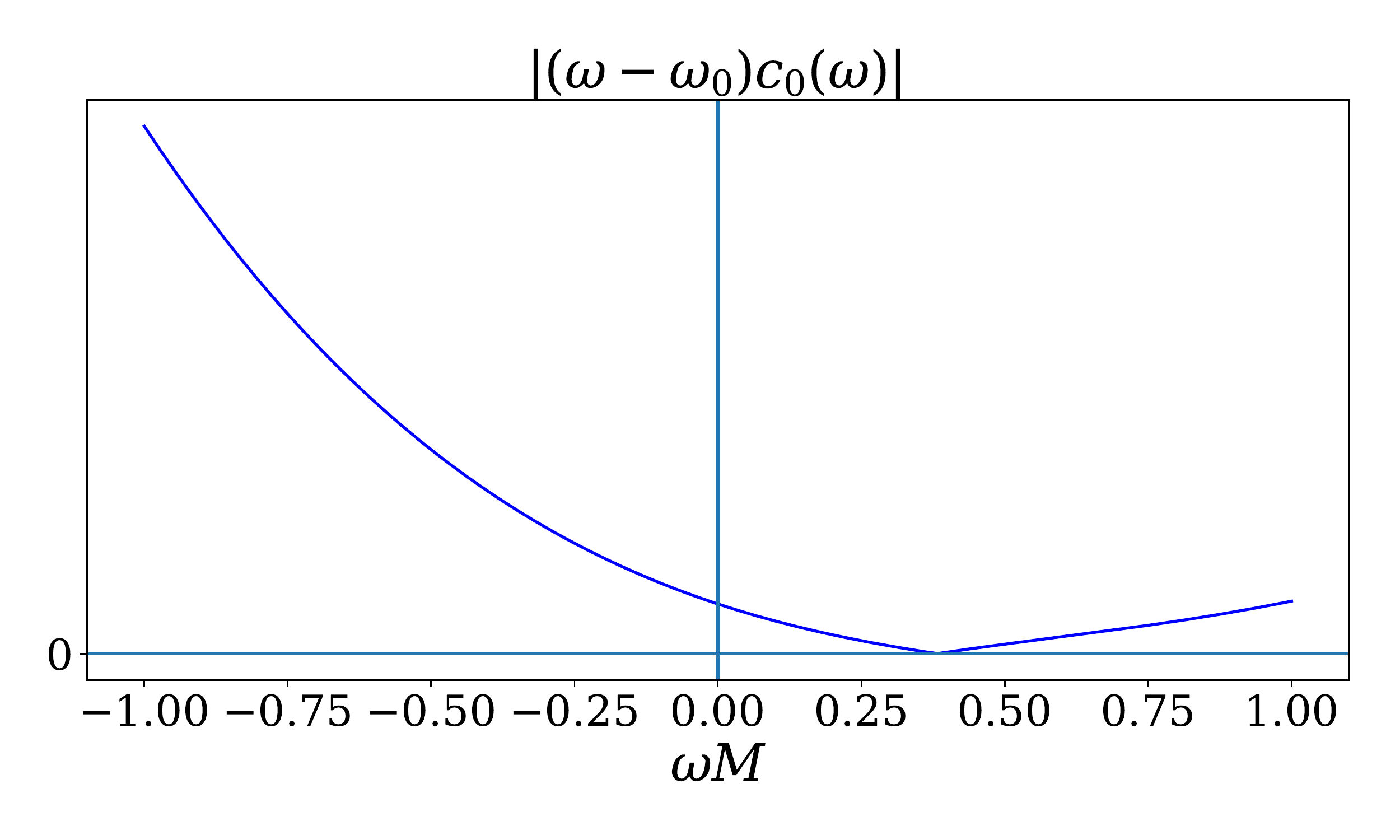} \\ 
 \includegraphics[width=0.49\textwidth]{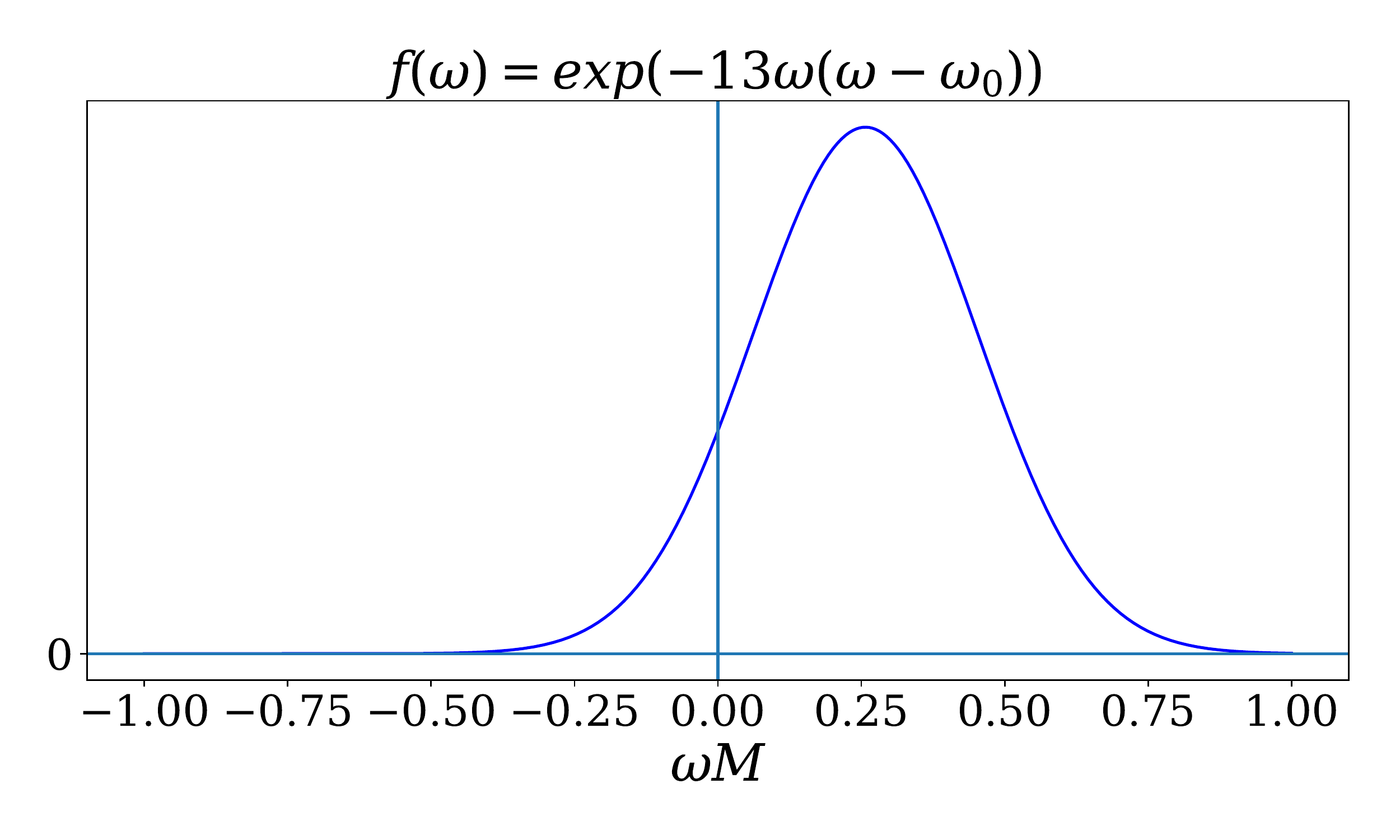}
 \includegraphics[width=0.49\textwidth]{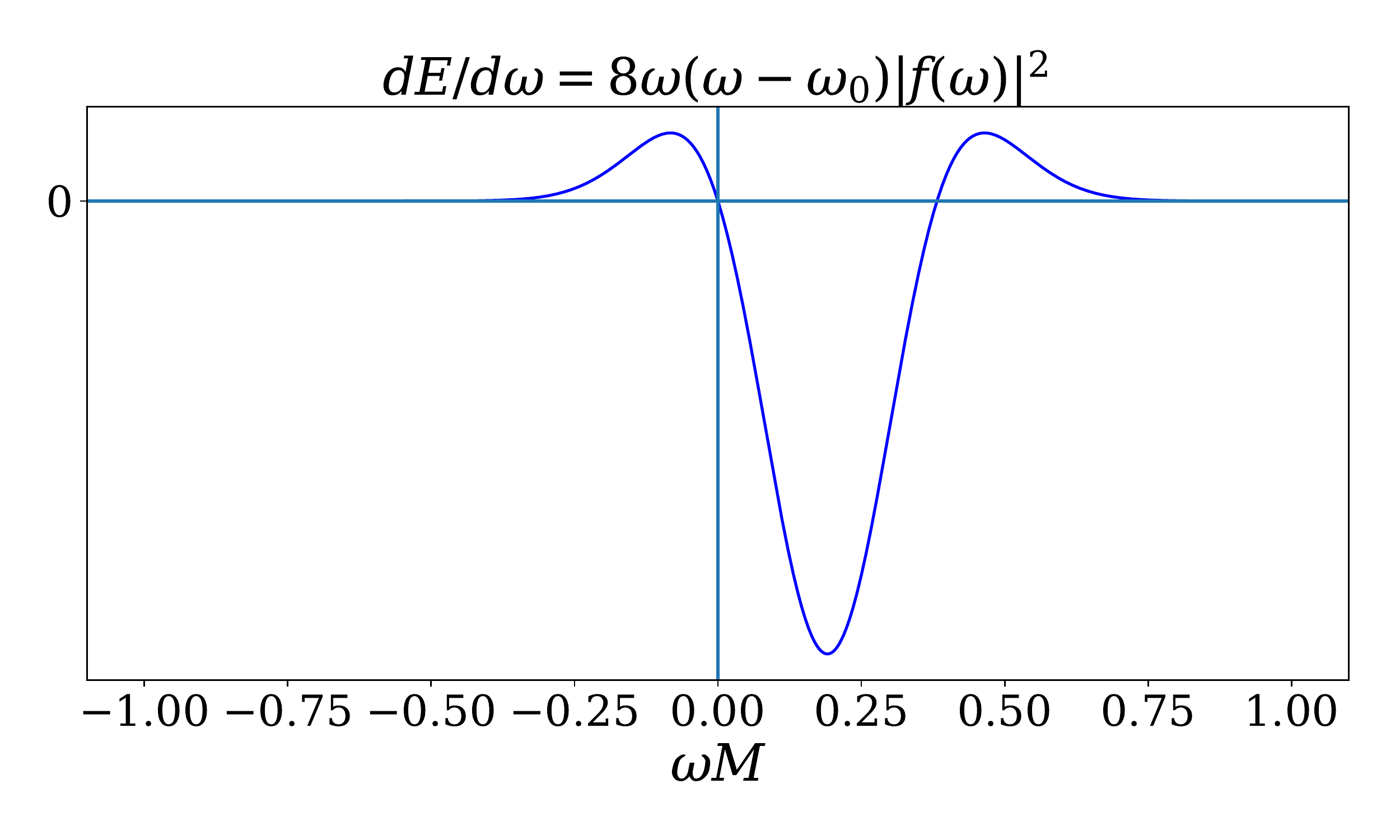}  
\caption{a) A universal factor appearing in the source integral $D(\omega)$. b) Function appearing in the initial pulse definition. c) The resulting spectral flux density.}
\label{fig1}
\end{figure}

We return to our result (\ref{e5}) for $D(\omega)$ in the case of inside pulses. $|D(\omega)|$ contains the factor $|(\omega-\omega_0)c_0(\omega)|$ which, as shown in Fig.~\ref{fig1}a, implies a vanishing signal as $\omega\rightarrow\omega_0$. This is the universal factor that substantially enhances negative frequencies relative to positive frequencies. For $f_{\leftarrow\atop \rightarrow}(\omega)$ we choose
\begin{align}
f_\rightarrow(\omega)=f_\leftarrow(\omega)=\rho e^{-\kappa\omega(\omega-\omega_0)},
\label{e7}\end{align}
a Gaussian centred at $\omega_0/2$. This will enhance positive vs negative frequencies and it will do so more for a thinner Gaussian, i.e.~larger $\kappa$. For the following we adopt $\kappa=13$, as shown in Fig.~\ref{fig1}b. Fig.~\ref{fig1}c displays the resulting flux density $dE/d\omega$, being negative inside the region $0<\omega<\omega_0$ and positive outside. The picture here is that the lowest energy modes, i.e.~the most negative ones, are excited the easiest, while modes with frequencies far away from $\omega_0/2$ are exponentially suppressed. Note that our initial conditions have the property that $dE/d\omega$ is continuous and is thus vanishing at $\omega=0$ and $\omega=\omega_0$; we do not consider a discontinuous $dE/d\omega$.

From (\ref{e5}) we see that with two or more travelling pulses, there will be interference effects that can cause $|D(\omega)|$ to oscillate. These oscillations can be rapid when pulses start far from the wall ($x_i-x_w$ is large). For example an ingoing and outgoing pulse both starting near the middle of the cavity will lead to an oscillation of $|D(\omega)|$ with a peak spacing that is close to twice the spacing between the resonance peaks in the transfer function. Thus the net effect is that every second spike in the absolute value of the final signal is suppressed. Moving the starting point of the pair of pulses away from the midpoint, in either direction, will produce a more slowly varying modulation of the resonance spikes. In particular, moving the pair of pulses close to the potential barrier means that the peak spacing in $|D(\omega)|$ almost matches the resonance spacing, and so the modulation is slowly varying.

In general two pulses starting at different positions and/or moving in opposite directions will generate two echo trains interleaved with each other. The time delay between one echo and the next can then be arbitrary, and only the time delay between every second echo is the standard $2\Delta x$. A larger number of pulses, each with a random $x_i$, direction and amplitude can easily generate a time waveform with no echo structure. And yet, as we shall see, the resonance structure in frequency space can still be observed.

\begin{figure}[h]
\centering
 \includegraphics[width=0.49\textwidth]{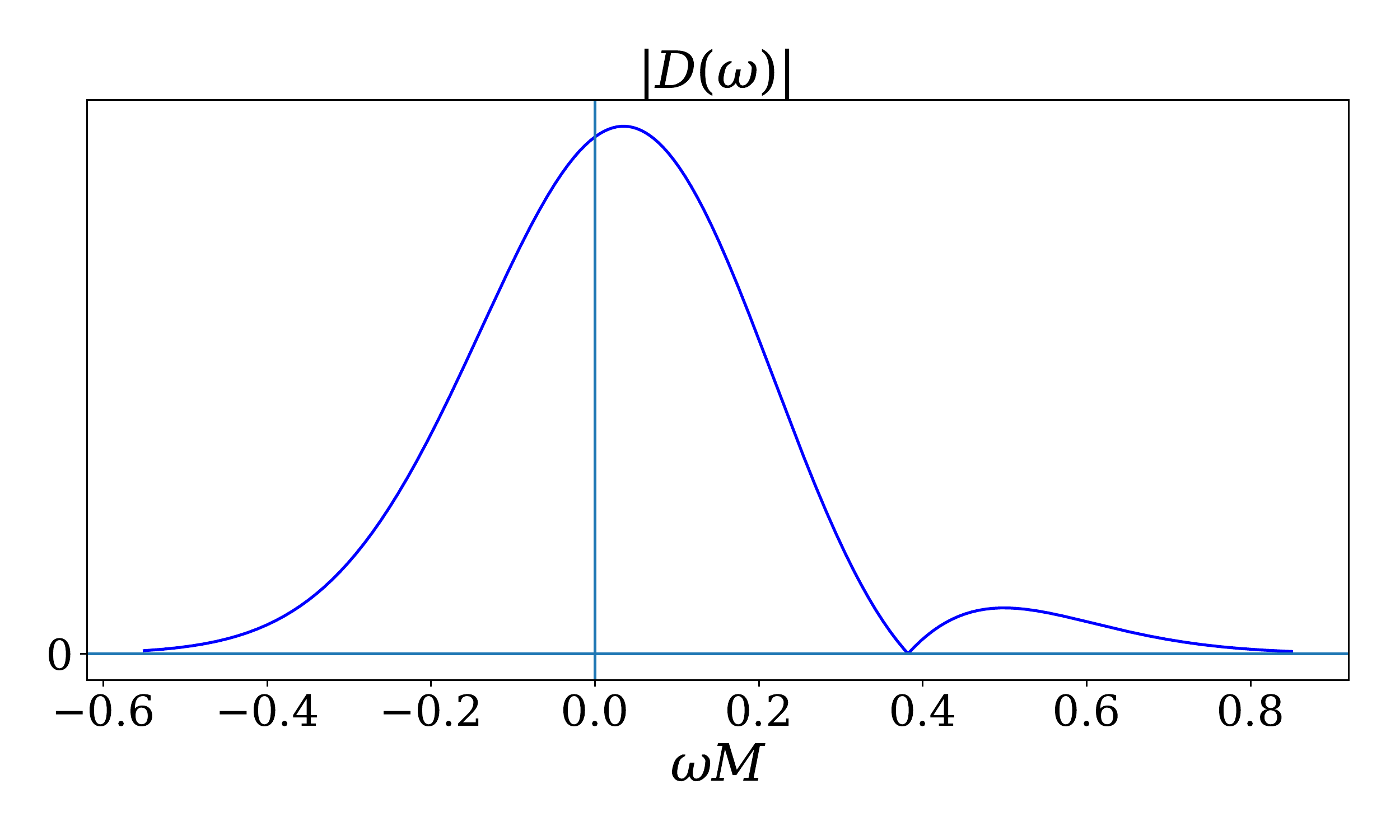}
 \includegraphics[width=0.49\textwidth]{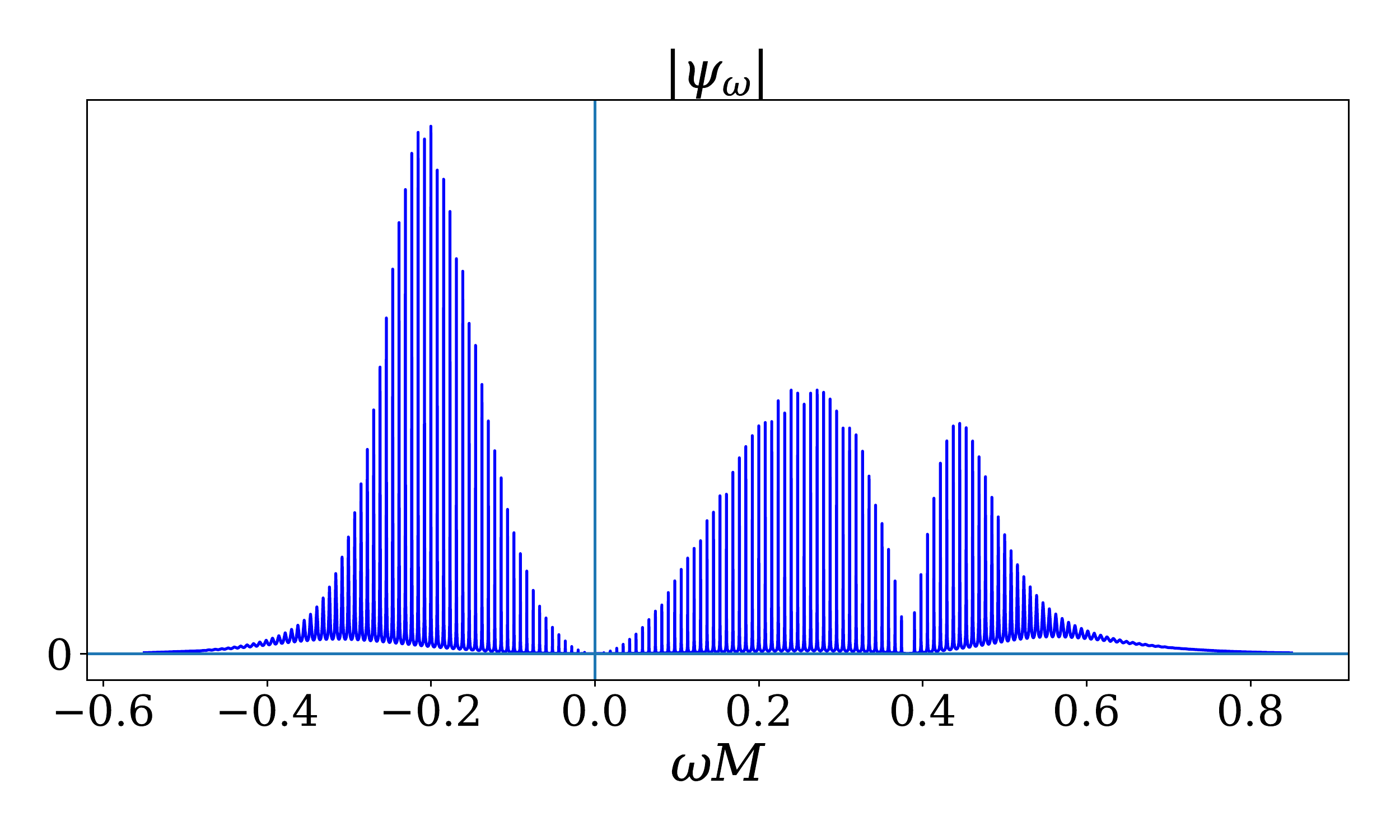}
 \includegraphics[width=0.49\textwidth]{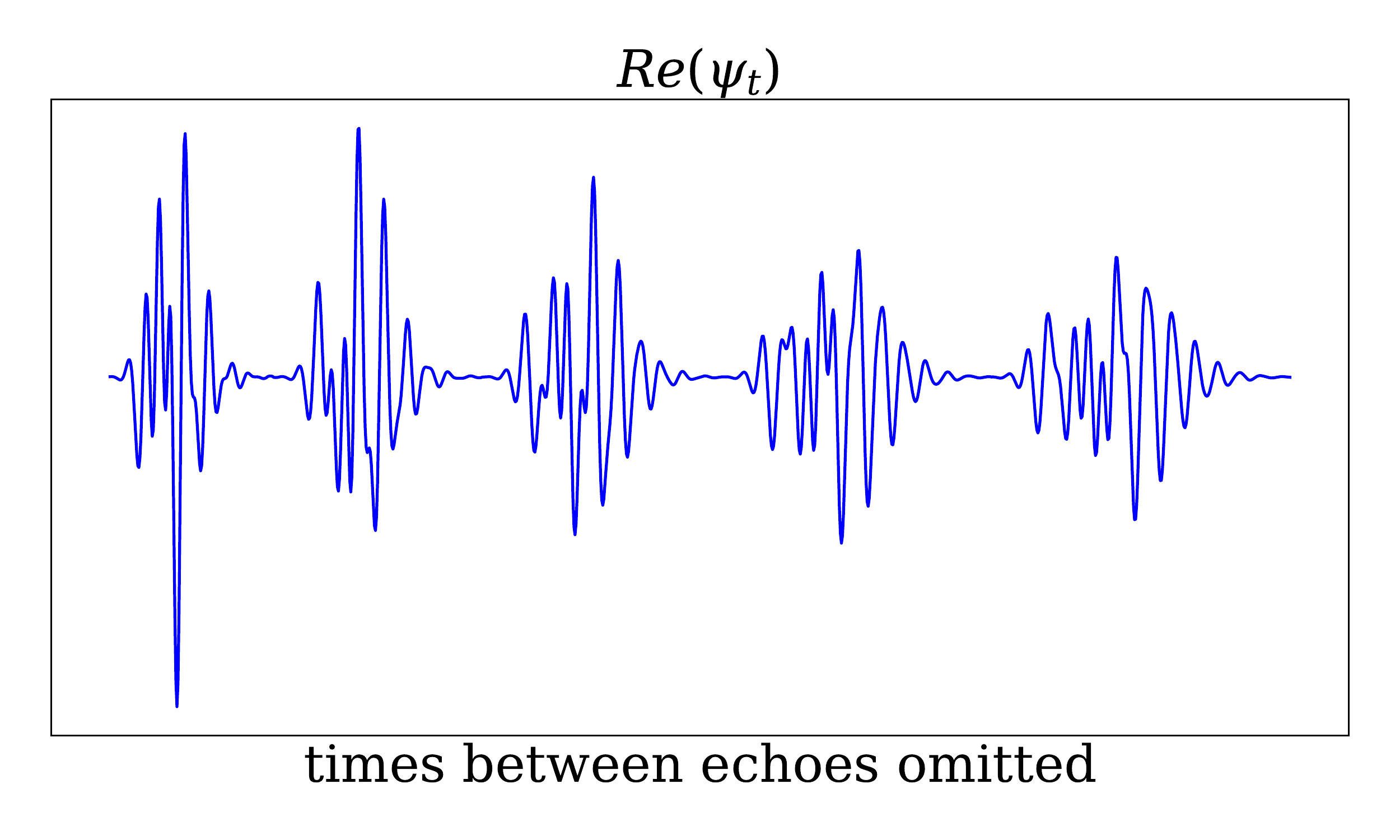}
 \includegraphics[width=0.49\textwidth]{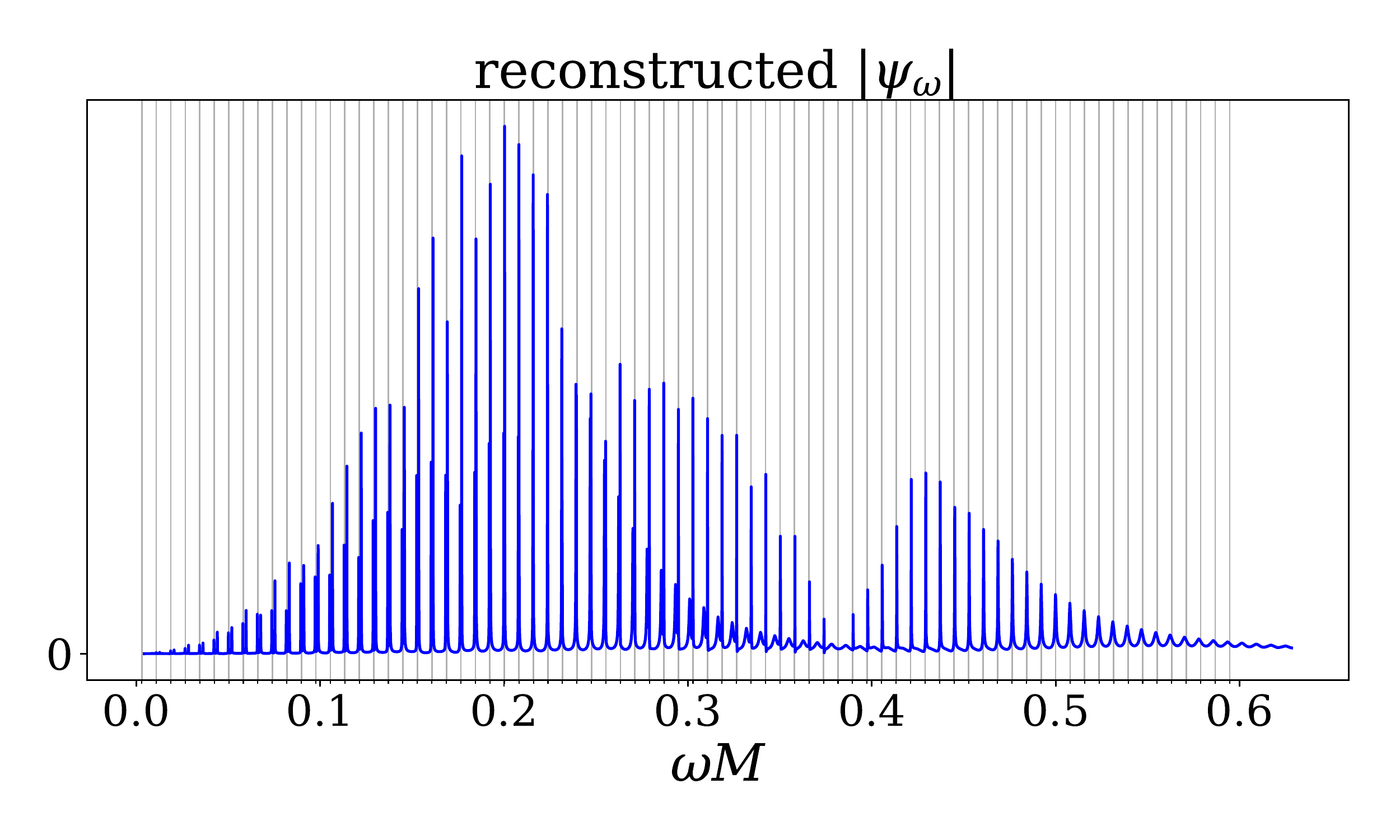}
\caption{Results for case 1. a) Absolute value of source integral. b) Absolute value of signal. c) First five echoes. d) Absolute value of reconstructed signal.}
\label{fig3}
\end{figure}
\begin{figure}[h]
\centering
 \includegraphics[width=0.49\textwidth]{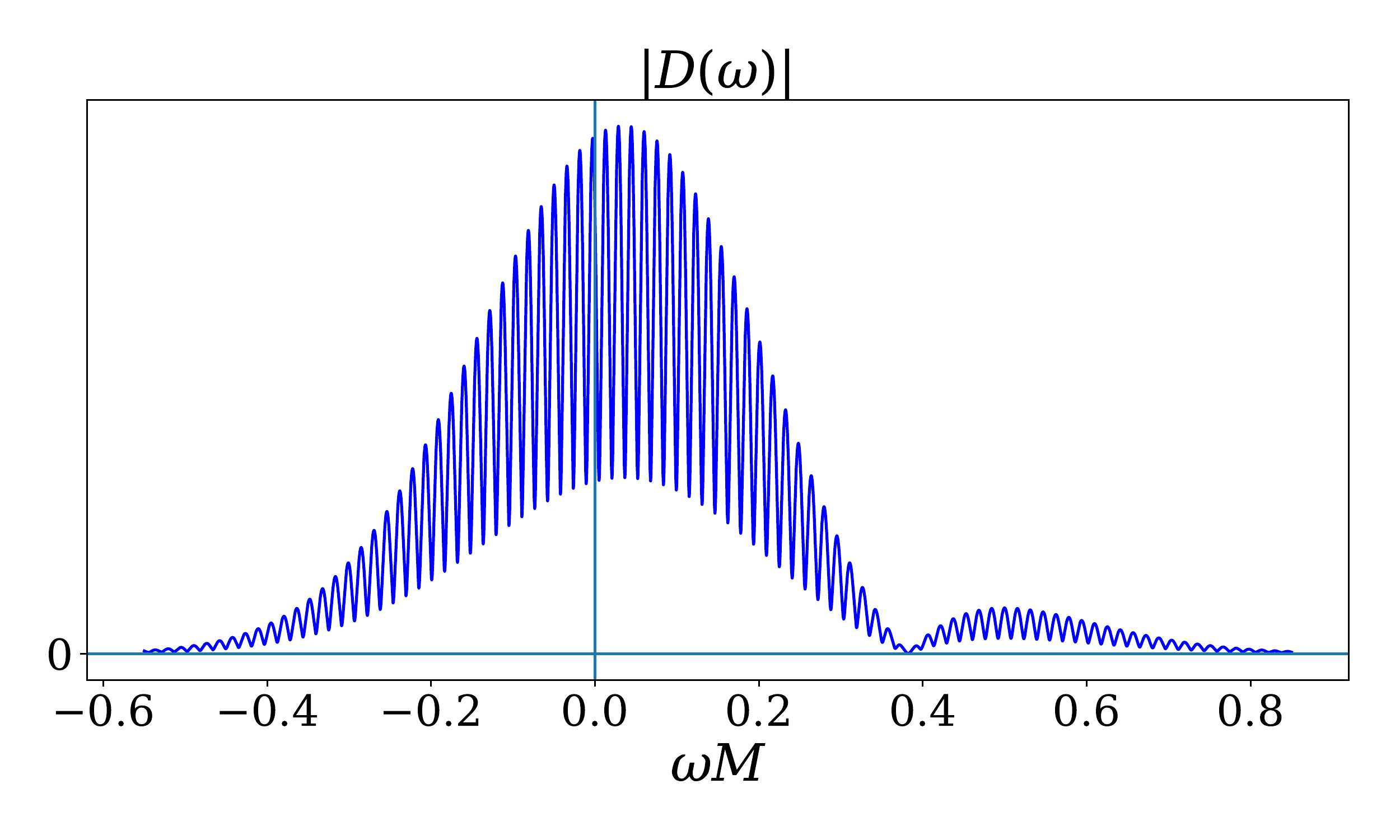}
 \includegraphics[width=0.49\textwidth]{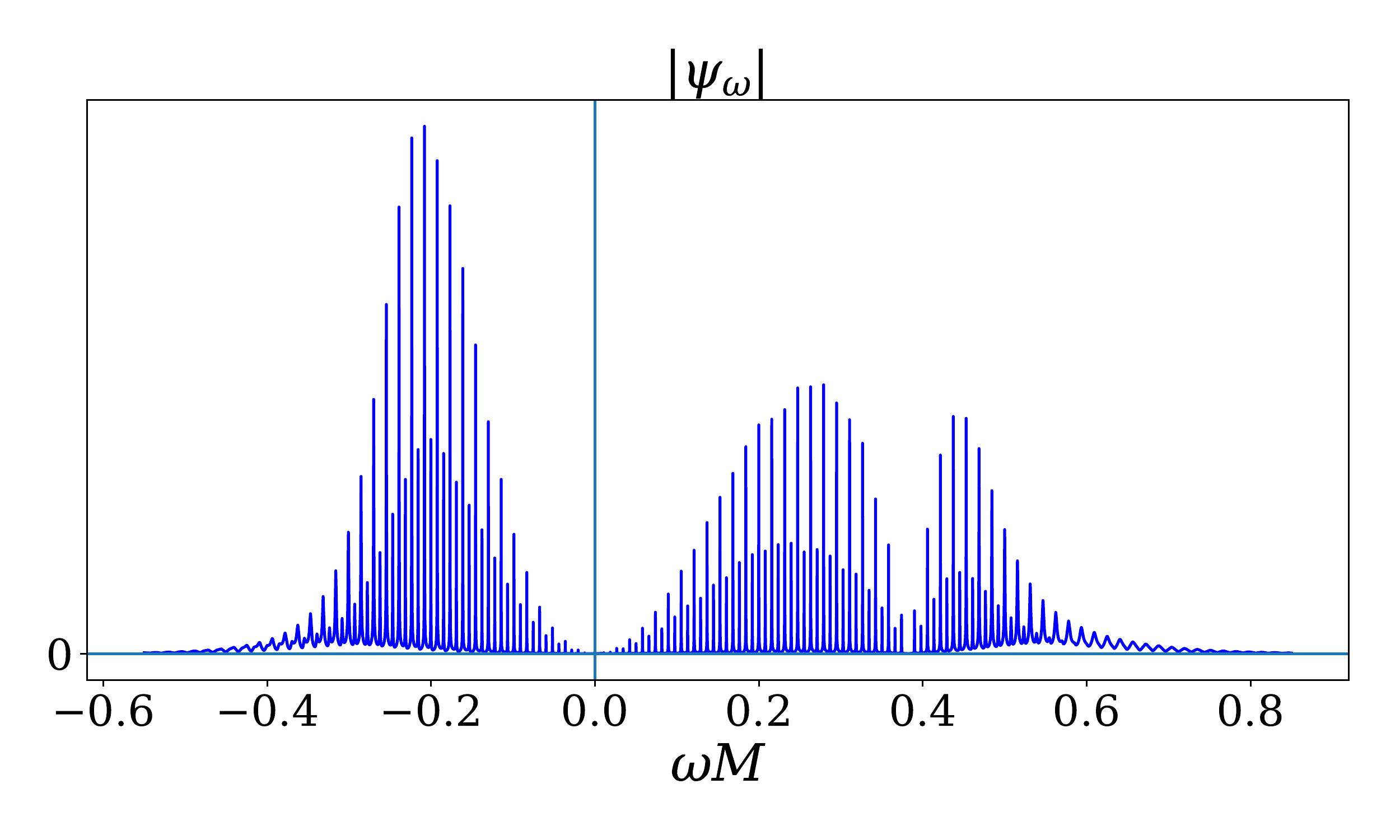}
 \includegraphics[width=0.49\textwidth]{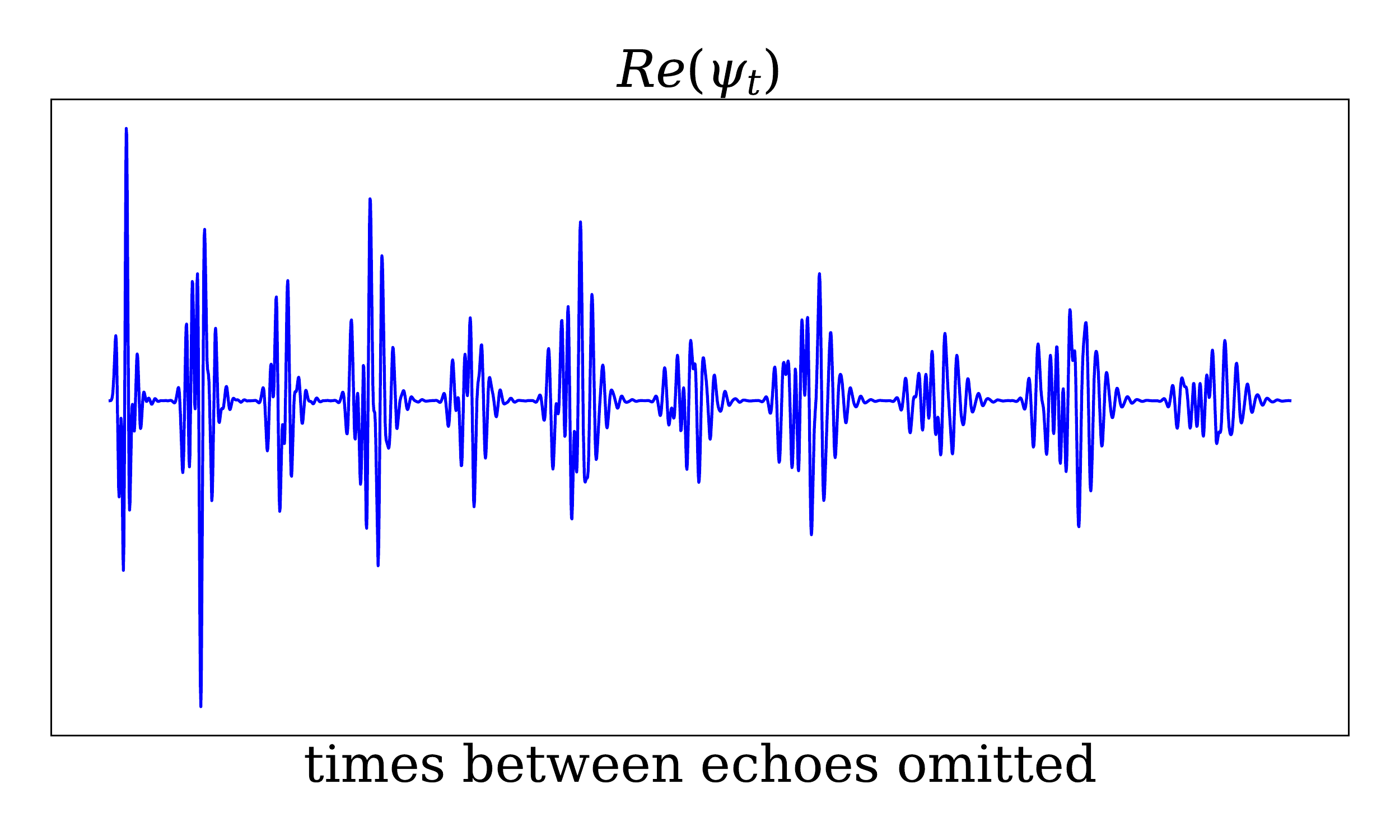}
 \includegraphics[width=0.49\textwidth]{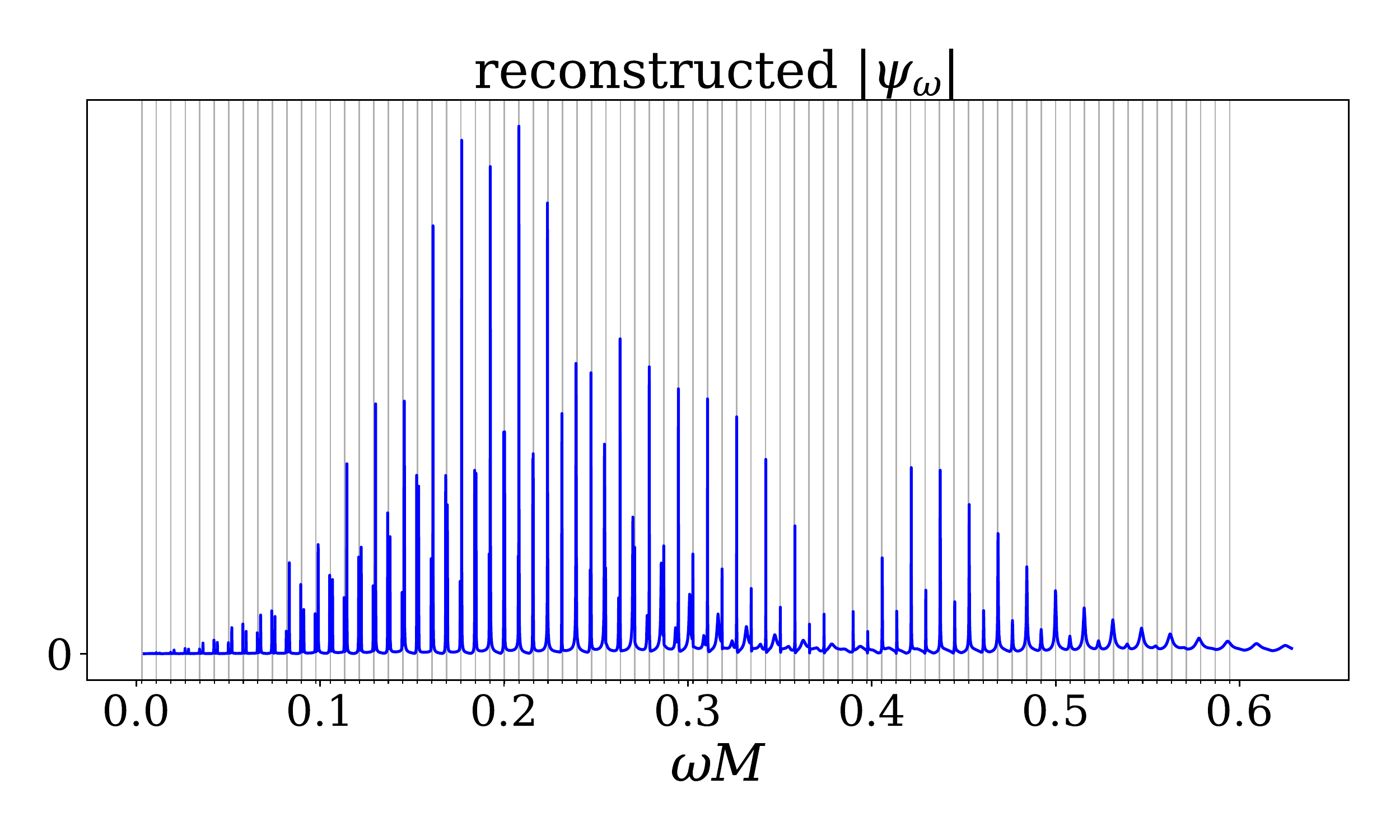}
\caption{Results for case 2.}
\label{fig4}
\end{figure}
\begin{figure}[h]
\centering
 \includegraphics[width=0.49\textwidth]{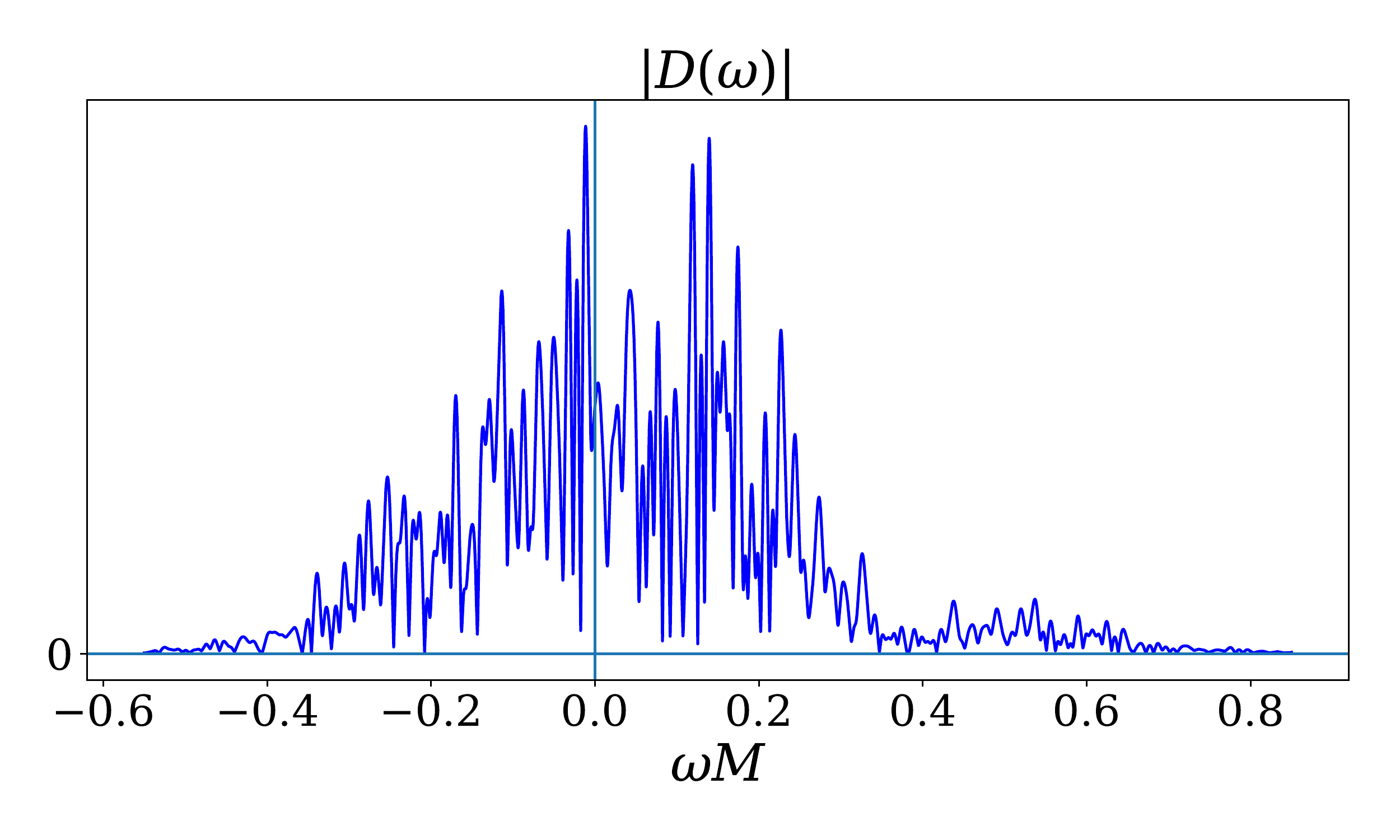}
 \includegraphics[width=0.49\textwidth]{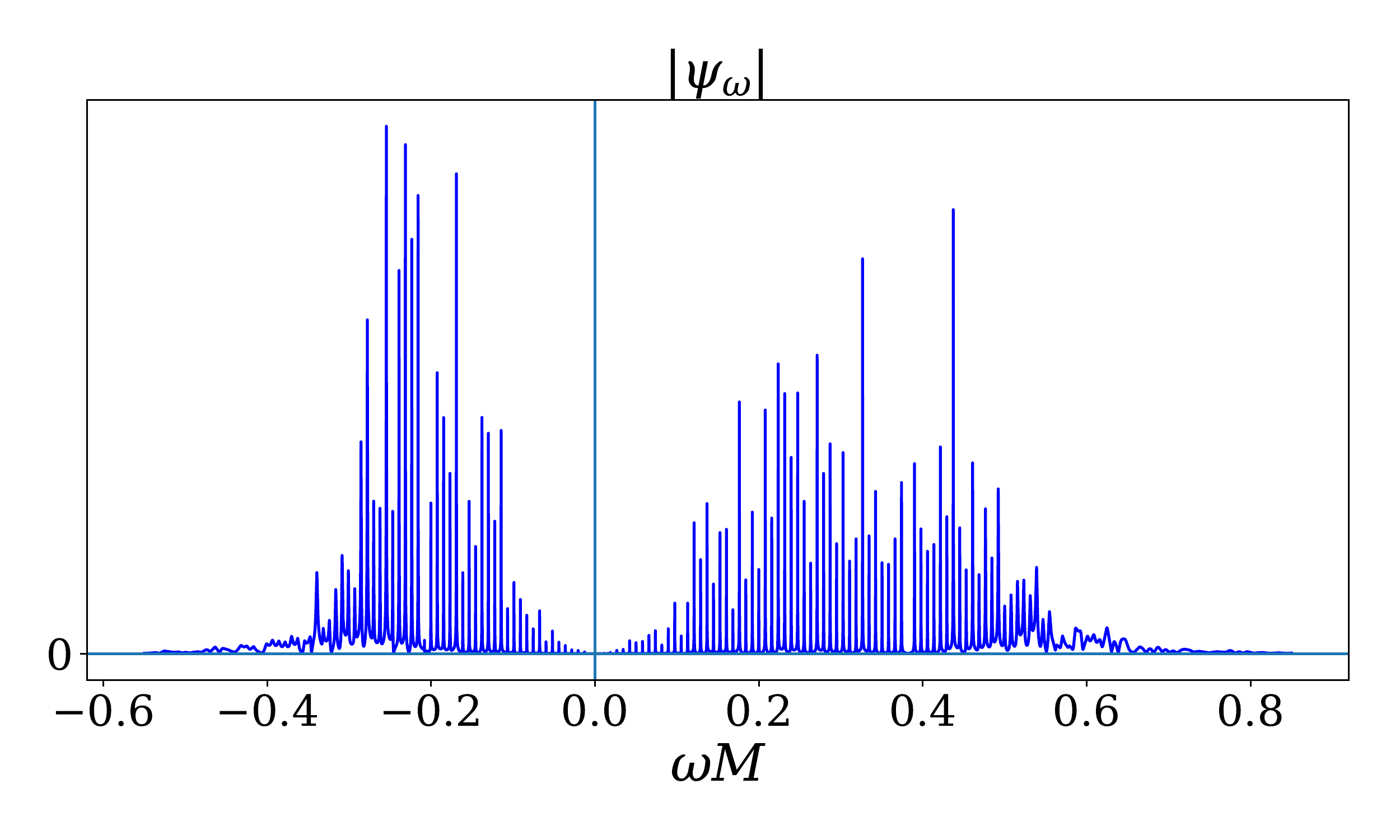}
 \includegraphics[width=0.49\textwidth]{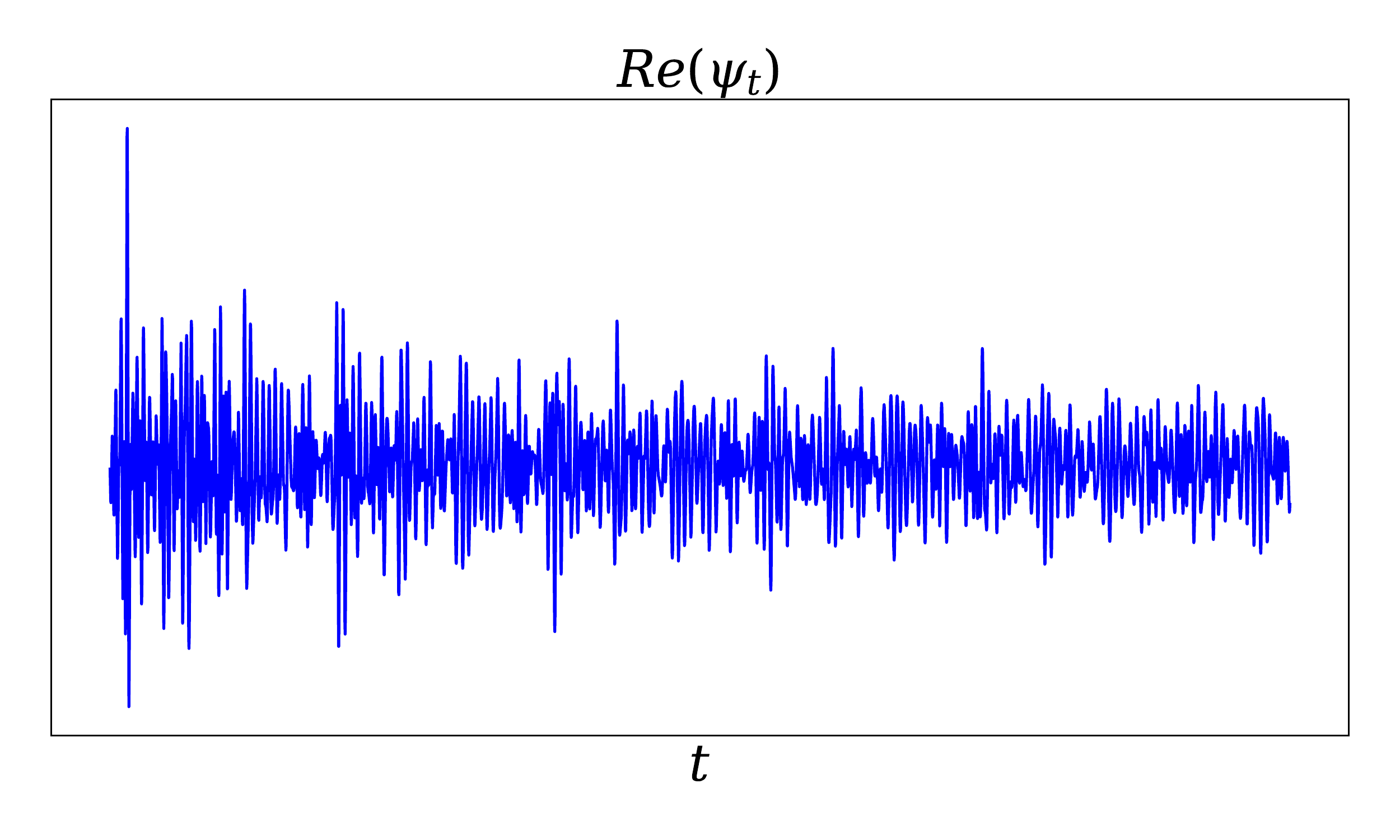}
 \includegraphics[width=0.49\textwidth]{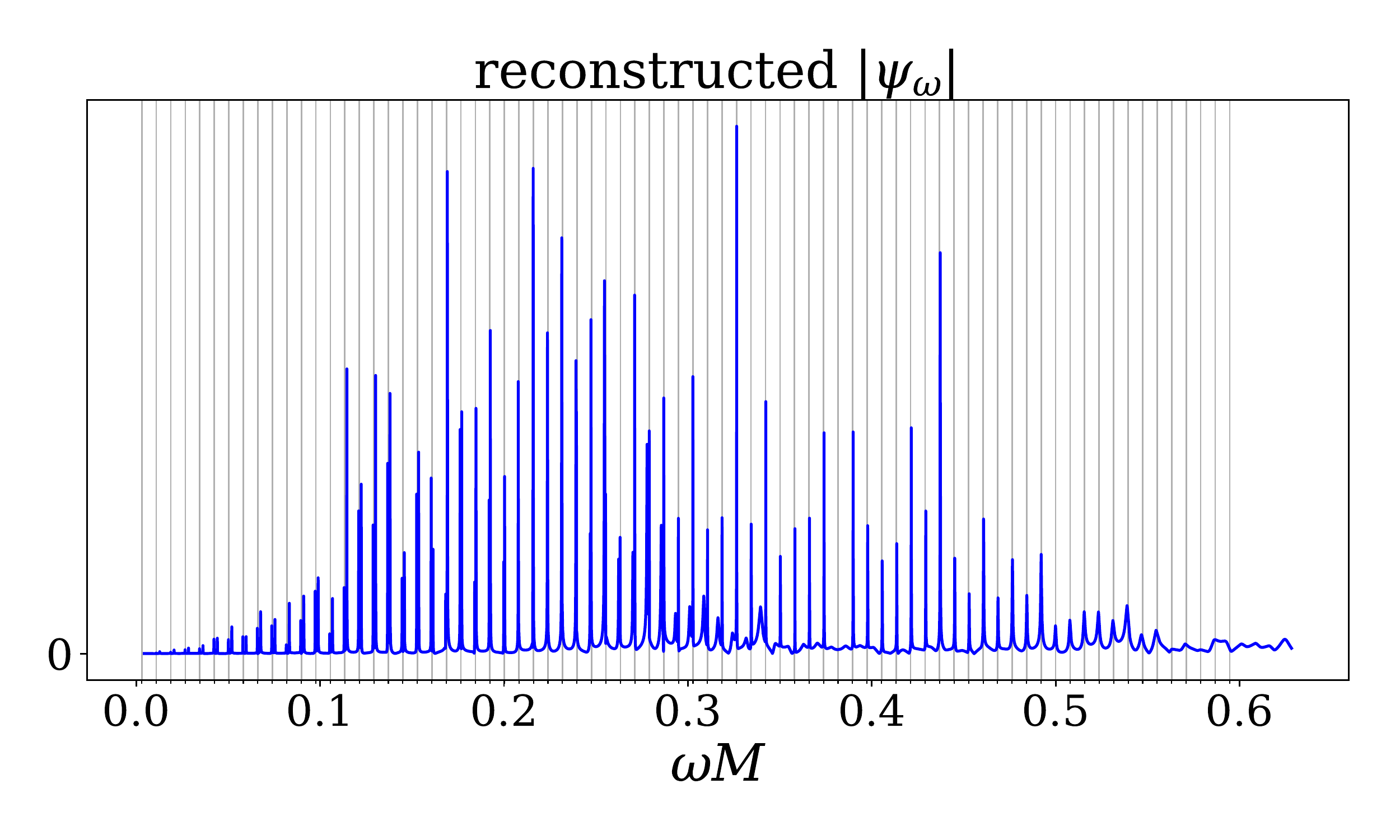}
\caption{Results for case 3.}
\label{fig5}
\end{figure}

In more detail, we consider the following choices for perturbations originating on the inside of the potential. The $f_{\leftarrow\atop \rightarrow}(\omega)$'s are chosen to be the Gaussian in (\ref{e7}).
\begin{enumerate}
\item A single pulse, can be either outgoing or ingoing,
\item an outgoing pulse and an ingoing pulse both starting at the centre of the cavity at $x_1=x_2=-200$,
\item 50 outgoing and 50 ingoing pulses, each with random position $x_i=(-400..0)$, random amplitude $\rho=(-1..1)$ and random narrowness of Gaussian $\kappa=(2..20)$.
\end{enumerate}

The results for case 1 are shown in Fig.~\ref{fig3}. Fig.~\ref{fig3}a shows that the source integral $|D(\omega)|$ peaks at $\omega=0$, which occurs due to our choice of the Gaussian with $\kappa=13$. Fig.~\ref{fig3}b shows the spectrum $|\psi_\omega|$, the product of the transfer function in Fig.~\ref{fig2} and the source integral. The additional dip around $\omega=\omega_0$ is the result of the $\omega-\omega_0$ factor in $D(\omega)$, and we see how the negative frequency spectrum is relatively enhanced as compared to the transfer function in Fig.~\ref{fig2}. Fig.~\ref{fig3}c shows the first five echoes in the real part of the time waveform. We see echoes with irregular shape, even for this simplest of initial conditions. This is not surprising given the three bump structure of $|\psi_\omega|$. 

Fig.~\ref{fig3}d shows the reconstructed spectrum as described above. As with all our examples, this is constructed from a time waveform with a duration of 180 echoes. We see how the positive and negative frequency spectra have been superimposed. As seen in Fig.~\ref{fig2}, the negative frequency spectrum is at lower values of $|\omega|$ than the positive frequency spectrum, and so the corresponding component of the reconstructed spectrum occurs at relatively low frequencies. It is also shown how a regularly spaced set of grid lines can still line up with most of the spikes.

Case 2 in Fig.~\ref{fig4} describes the case where the initial state in the cavity is described by a perturbation at the potential barrier and another perturbation at the wall. This is achieved by starting an ingoing and an outgoing pulse at the centre of the cavity, so that they reach the two ends of the cavity simultaneously. We also choose the amplitude of the ingoing pulse to be half that of the outgoing pulse. Fig.~\ref{fig4}a shows how $|D(\omega)|$ rapidly oscillates, and the result is to suppress every second spike in the resulting $|\psi_\omega|$. The reconstructed spectrum $|\psi_\omega|$ also shows this effect. In the time domain the result is to have a set of echoes with half the amplitude interspersed half-way between the original set of echoes.

Case 3 in Fig.~\ref{fig5} shows the case of many random pulses populating the whole cavity. $|D(\omega)|$ is now very noisy and any simple echo structure in the time waveform is gone. But evenly spaced spikes in $|\psi_\omega|$ and its reconstructed version are still present. If not already obvious, this shows the superiority of a search strategy that is focused on the evenly spaced structure in the frequency data.
\begin{figure}[h]
\centering
 \includegraphics[width=0.49\textwidth]{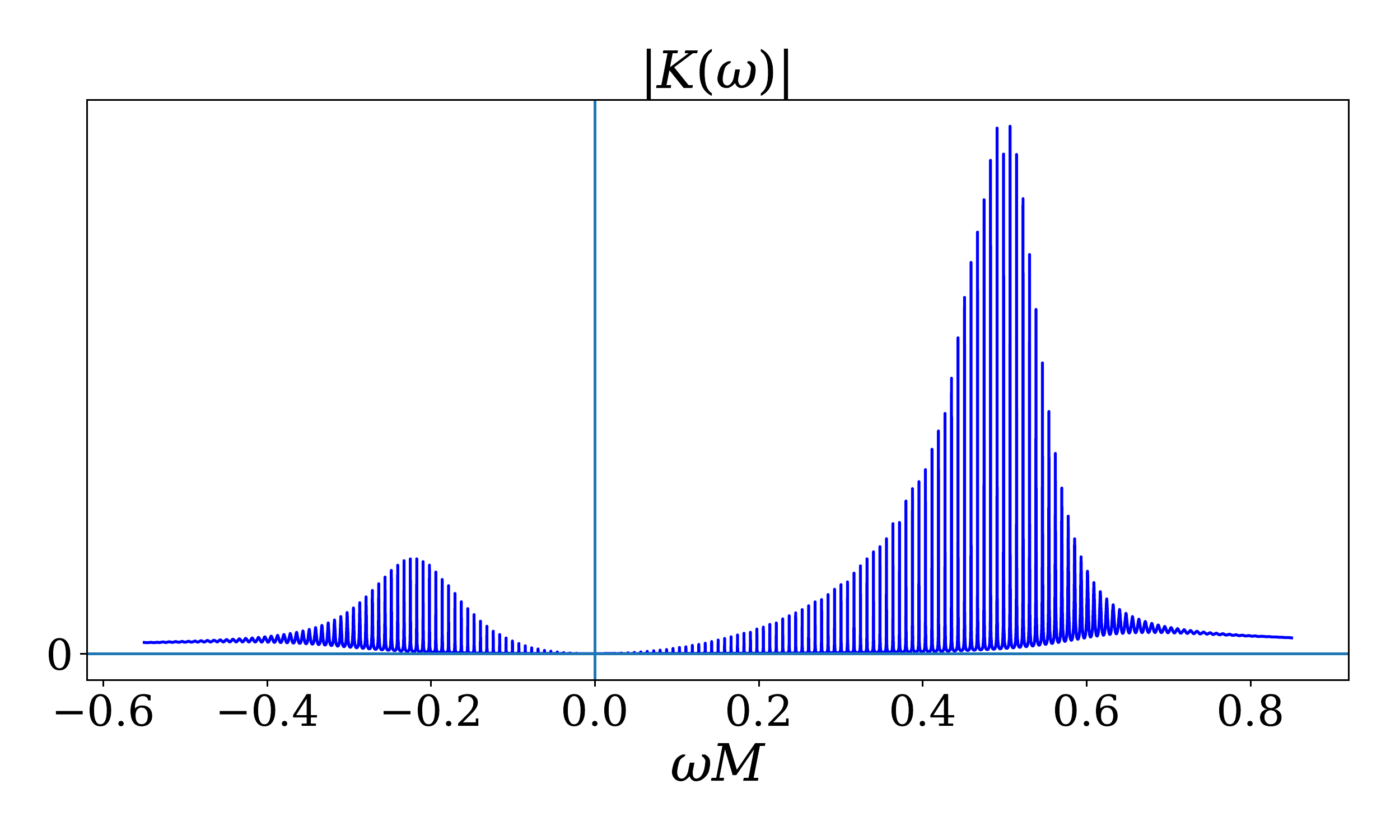} 
 \includegraphics[width=0.49\textwidth]{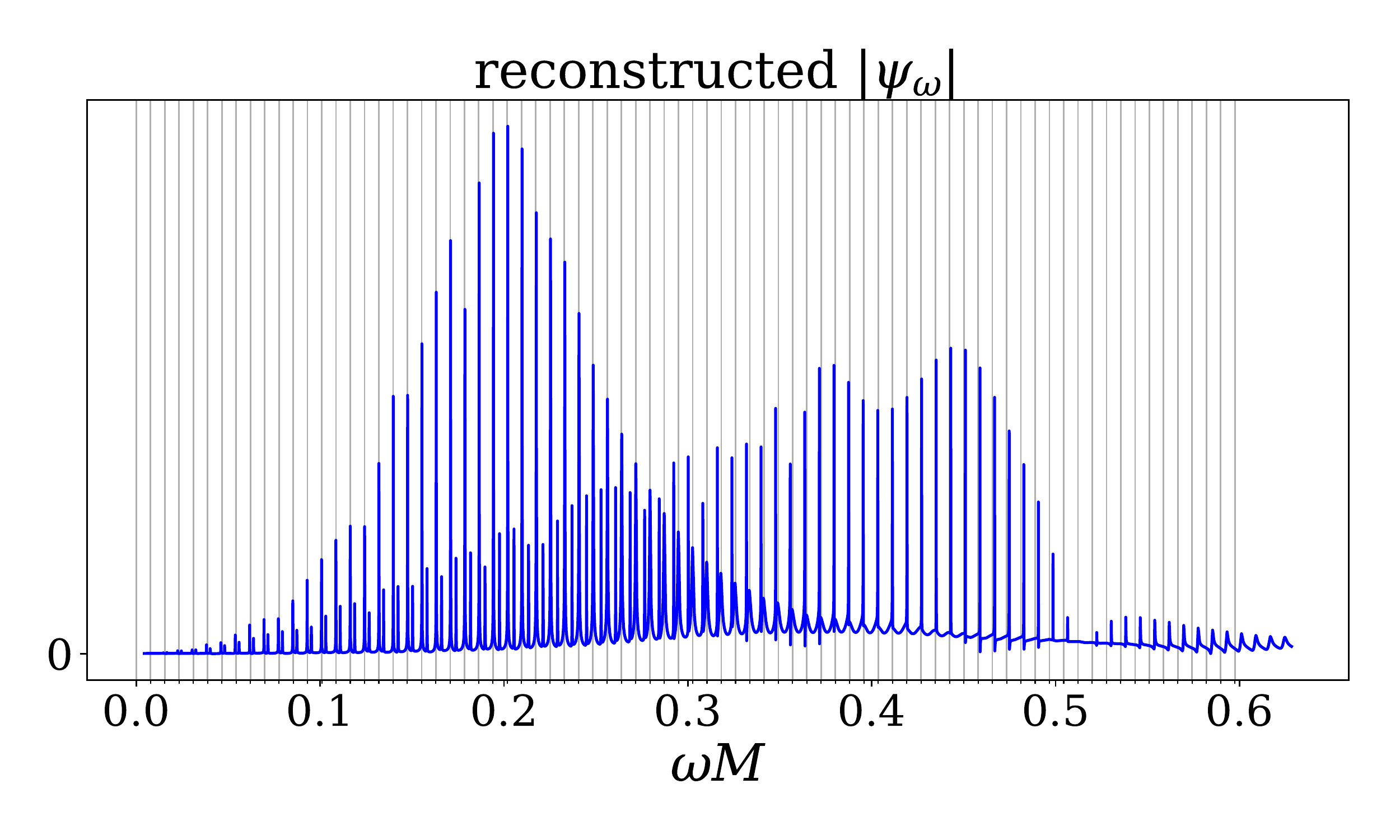}
\caption{Results for spin $\chi=.81355$ for case 1.}
\label{fig7}
\end{figure}

All of the above has been for a spin of $\chi=2/3$, but we briefly consider a higher spin of $\chi=0.81355$.\footnote{This special value of spin causes $b_0(\omega)$ to have a zero at $\omega\approx 0.90$. This causes no particular problem for us because of our energy normalization of the pulses.} In one of the LIGO merger events (GW170729) a final black hole spin is reported close to this value. For this spin we have the value $\omega_0=0.5144$. For our Gaussian description of $f_{\leftarrow\atop \rightarrow}(\omega)$ we now choose $\kappa=5$. Then many of the plots look qualitatively similar to the $\chi=2/3$ case after accounting for the change in $\omega_0$. We provide two plots in Fig.~\ref{fig7}, showing the transfer function and the final reconstructed spectrum for case 1. From Fig.~\ref{fig7}a we see that the higher spin causes the positive frequency component to move further away from $\omega=0$, and so in Fig.~\ref{fig7}b the two components are further apart. It is also seen that a sharp resonance spectrum now extends over a broader range of frequencies, meaning that a higher spin object has a stronger resonance signal.

At the end of the last section we discussed the case of perturbations impinging on the potential barrier from the outside. We described how $\psi_\omega$ is then described by a completely different structure in the complex plane, where every pole is now mirrored by a zero on the other side of the real $\omega$ axis. But the poles in the upper half-plane, in particular, are still present, and this again requires an inverse Laplace transform to obtain the waveform in the time domain. There is again the $i\epsilon$ shift of the integration contour, and because of this the effects of the poles are no longer invisible, even in the case of perfect reflection. The new feature is that the amplitude of each QNM is suppressed by the separation between each pole and its corresponding zero, or in other words by the imaginary part of the pole position. This is very small for all the very sharp resonances, in particular those in the superradiance region. Thus the unstable modes in this case are hardly excited. The QNMs that now dominate are those that decay more quickly.

From this it is clear why the echo waveform that is generated by a perturbation on the outside has echo amplitudes that fall more rapidly than for echoes generated by an inside perturbation. And in frequency space, the striking resonance spectrum that we have been discussing for the inside perturbation is almost completely eliminated for the outside perturbation. Thus if signals that we have described in this section are seen then we immediately know that we are seeing the consequences of initial perturbations \textit{inside} the cavity.

\section{Conclusion}

We have described how the spectral content of an echo waveform is the product of a transfer function and an initial-condition-dependent source integral. The closely spaced resonance spectrum becomes modulated. The modulation typically has a three bump structure as shown in Figs.~\ref{fig3}b and \ref{fig4}b, with one bump at negative frequencies and two at positive frequencies, with the latter being due to the source integral having a zero at $\omega_0$. Because of this generic structure the echo waveforms in the time domain have a more nontrivial structure than has been considered before, with the complex shape of each echo changing from one to the next. We have also considered the effect of the experimental projection of the two polarizations into the real strain data, and we expressed this effect in terms of a reconstructed spectrum, as in Figs.~\ref{fig3}d and \ref{fig4}d. The reconstructed spectra also depends on the time duration, taken here to contain 180 echoes.

As we have varied the form of the initial perturbation we have seen that the structure of the echoes, and whether they even exist, is strongly affected. In Figures (\ref{fig3}c-\ref{fig5}c) there are three examples of waveforms in the time domain that are dramatically different from each other. Thus in addition to the generic complexity, there is additional nontrivial and difficult-to-model information in the waveform that is sensitive to initial conditions. In Fig.~\ref{fig5}d we see how the evenly spaced spike pattern in the reconstructed spectrum $|\psi_\omega|$ persists even when distinct echoes are completely absent. The obvious conclusion is that a resonance search is preferred over a search that tries to identify the full echo waveform, as in a matched-filter analysis.

Other than \cite{Conklin:2017lwb}, all other echo searches have been based on matched-filter analyses. This began with \cite{Abedi:2016hgu} and the template developed there has also become the basis for other analyses \cite{Westerweck:2017hus,Lo:2018sep,Nielsen:2018lkf,Uchikata:2019frs}. We see that our simplest waveform in Fig.~\ref{fig3}c is significantly more complex than that template, meaning that these particular analyses would have trouble identifying an injection of even our simplest signal into noise, let alone injections of our more complex waveforms. In \cite{Tsang:2019zra} templates are constructed from a superposition of some number of nine-parameter generalized sine-Gaussians. The large parameter space means more false positives from noise and thus a sensitivity that decreases with the square root of the number of parameters. Such an approach would need to be restricted to a small number of echoes.

The reconstructed spectrum of interest for a resonance search is also seen to have additional structure. There is an interleaving of two resonance structures, which is traced back to the positive and negative frequency resonances of the original transfer function. We have found natural enhancement factors for the negative frequency component from the source integral. This enhancement may be offset to some extent by energy considerations that could imply larger amplitudes for the positive frequency modes in the initial perturbed state. Thus observing the relative size of the positive and negative energy components will tell us about the initial state of the newly formed not quite black hole.

And finally we have found that an evenly spaced comb can still line up with the frequency spikes at both high and low frequencies, even though they are coming from the two different components. Looking at our reconstructed $|\psi_\omega|$ plots carefully, though, shows that the comb does not quite line up with spikes at intermediate frequencies. The misalignment is greater for our higher spin example. There may be some cases where a comb that only attempts to match one component could give a larger signal.

\appendix*
\section{Amplitude relations}
In the asymptotic regions the Sasaki and Nakamura amplitudes take the simple form 
\begin{eqnarray}\label{eq:SNEamp}
X_{lm\omega}\to
\left\{\begin{array}{cc}
A_\textrm{trans}e^{-i k_H x}+A_\textrm{ref}e^{i k_H x},& x\to-\infty\\
A_\textrm{in}e^{-i \omega x}+A_\textrm{out}e^{i \omega x},& x\to\infty,\end{array}
\right.
\end{eqnarray}
while the Teukolsky amplitudes are
\begin{eqnarray}\label{eq:TEamp}
R_{lm\omega}\to
\left\{\begin{array}{cc}
B_\textrm{trans}\Delta^2 e^{-i k_H x}+B_\textrm{ref}e^{i k_H x},& x\to-\infty,\\
B_\textrm{in}\frac{1}{r}e^{-i \omega x}+B_\textrm{out}r^3 e^{i \omega x},& x\to\infty.\end{array}
\right.
\end{eqnarray}
Here we use the notation $k_H=\omega-\omega_0$. The relations between these amplitudes are 
\begin{eqnarray}
B_\textrm{in}=-\frac{1}{4 \omega ^2}A_\textrm{in},\quad
B_\textrm{out}=-\frac{4 \omega ^2}{c_0}A_\textrm{out},\quad
B_\textrm{trans}=\frac{1}{d}A_\textrm{trans},\quad
B_\textrm{ref}=\frac{1}{g}A_\textrm{ref}.
\end{eqnarray}
The first three are given in~\cite{Sasaki:2003xr} and the fourth was obtained in \cite{Conklin:2017lwb}. The various coefficients are
\begin{align}
c_0&=\lambda(\lambda+2)-12a\omega(a\omega-m)-i 12\omega M,\nonumber\\
d&=-4(2M r_+)^{5/2}\left[(k_H^2-8\epsilon^2)+i 6 k_H \epsilon\right],\nonumber\\
g&=\frac{-b_0}{4k_H(2Mr_+)^{3/2}(k_H+i 2\epsilon)},\\
b_0&=\lambda ^2+2 \lambda-96 k_H^2 M^2+72 k_H M r_+ \omega -12 r_+^2 \omega ^2\nonumber\\&-i [16 k_H M \left(\lambda+3-3\frac{M}{r_+}\right)-12 M \omega -8 \lambda  r_+ \omega],\nonumber\\
\epsilon&=(r_+-M)/(4M r_+).
\end{align}
where $\lambda$ is the spheroidal harmonic eigenvalue of the Teukolsky angular equation. 

The spectral flux densities $dE/d\omega$ at the horizon and at spatial infinity are expressed in terms of Teukolsky amplitudes in \cite{Brito:2015oca, Nakano:2017fvh}. We can use the conversions above to obtain
\begin{equation}
\frac{d E_\textrm{out}}{d\omega} = \frac{8 \omega ^2}{|c_0|^2} |A_\textrm{out}|^2, \quad 
\frac{d E_\textrm{in}}{d\omega} = \frac{8 \omega^2}{|C|^2}|A_\textrm{in}|^2,\quad
\frac{d E_\textrm{trans}}{d\omega} = \frac{8 \omega k_H}{|C|^2} |A_\textrm{trans}|^2,\quad
\frac{d E_\textrm{ref}}{d\omega} = \frac{8 \omega k_H}{|b_0|^2} |A_\textrm{ref}|^2,
\end{equation}
where 
\begin{align}
|C|^2 = & \lambda ^4+4 \lambda ^3+\lambda ^2 \left(-40 a^2 \omega ^2+40 a m \omega +4\right)+48 a \lambda  \omega  (a \omega +m)\nonumber\\
& + 144 \omega ^2 \left(a^4 \omega ^2-2 a^3 m \omega +a^2 m^2+M^2\right) .
\end{align}

\begin{acknowledgements}
We thank Jing Ren for earlier collaborations and for her comments on this work. This research was supported in part by the Natural Sciences and Engineering Research Council of Canada.
\end{acknowledgements}


\begin{thebibliography}{99}

\bibitem{Holdom:2016nek} 
B.~Holdom and J.~Ren,
``Not quite a black hole,''
Phys.\ Rev.\ D {\bf 95}, no.~8, 084034 (2017)
[arXiv:1612.04889 [gr-qc]].

\bibitem{Berti:2009kk}
E.~Berti, V.~Cardoso and A.~O.~Starinets,
``Quasinormal modes of black holes and black branes,''
Class.\ Quant.\ Grav.\  {\bf 26}, 163001 (2009)
[arXiv:0905.2975 [gr-qc]].

\bibitem{Cardoso:2016rao} 
  V.~Cardoso, E.~Franzin and P.~Pani,
  ``Is the gravitational-wave ringdown a probe of the event horizon?,''
  Phys.\ Rev.\ Lett.\  {\bf 116}, no.~17, 171101 (2016)
  Erratum: [Phys.\ Rev.\ Lett.\  {\bf 117}, no.~8, 089902 (2016)]
  [arXiv:1602.07309 [gr-qc]].

\bibitem{Cardoso:2016oxy} 
V.~Cardoso, S.~Hopper, C.~F.~B.~Macedo, C.~Palenzuela and P.~Pani,
``Gravitational-wave signatures of exotic compact objects and of quantum corrections at the horizon scale,''
Phys.\ Rev.\ D {\bf 94}, no.~8, 084031 (2016)
[arXiv:1608.08637 [gr-qc]].

\bibitem{Conklin:2017lwb} 
R.~S.~Conklin, B.~Holdom and J.~Ren,
``Gravitational wave echoes through new windows,''
Phys.\ Rev.\ D {\bf 98}, no.~4, 044021 (2018)
[arXiv:1712.06517 [gr-qc]].

\bibitem{Mark:2017dnq} 
  Z.~Mark, A.~Zimmerman, S.~M.~Du and Y.~Chen,
  ``A recipe for echoes from exotic compact objects,''
  Phys.\ Rev.\ D {\bf 96}, no.~8, 084002 (2017)
  [arXiv:1706.06155 [gr-qc]].

\bibitem{Teuk}
S.~A.~Teukolsky, ``Rotating Black Holes: Separable Wave Equations for Gravitational and Electromagnetic Perturbations,'' Phys.~Rev.~Lett.~29, 1114 (1972).

\bibitem{Sasaki:1981sx}
M.~Sasaki and T.~Nakamura,
``Gravitational Radiation From a Kerr Black Hole. 1. Formulation and a Method for Numerical Analysis,''
Prog.\ Theor.\ Phys.\  {\bf 67}, 1788 (1982).

\bibitem{Nakano:2017fvh} 
H.~Nakano, N.~Sago, H.~Tagoshi and T.~Tanaka,
``Black hole ringdown echoes and howls,''
PTEP {\bf 2017}, no.~7, 071E01 (2017)
[arXiv:1704.07175 [gr-qc]].

\bibitem{Brito:2015oca} 
R.~Brito, V.~Cardoso and P.~Pani; ``Superradiance : Energy Extraction, Black-Hole Bombs and Implications for Astrophysics and Particle Physics,''
Lect.\ Notes Phys.\  {\bf 906}, pp.1 (2015)
[arXiv:1501.06570 [gr-qc]].

\bibitem{Maggio:2017ivp} 
E.~Maggio, P.~Pani and V.~Ferrari,
``Exotic Compact Objects and How to Quench their Ergoregion Instability,''
Phys.\ Rev.\ D {\bf 96}, no.~10, 104047 (2017)
[arXiv:1703.03696 [gr-qc]].

\bibitem{Maggio:2018ivz} 
  E.~Maggio, V.~Cardoso, S.~R.~Dolan and P.~Pani,
  ``Ergoregion instability of exotic compact objects: electromagnetic and gravitational perturbations and the role of absorption,''
  Phys.\ Rev.\ D {\bf 99}, no.~6, 064007 (2019)
 [arXiv:1807.08840 [gr-qc]].

\bibitem{Sasaki:2003xr} 
M.~Sasaki and H.~Tagoshi,
``Analytic black hole perturbation approach to gravitational radiation,''
Living Rev.\ Rel.\  {\bf 6}, 6 (2003)
[gr-qc/0306120].

\bibitem{Abedi:2016hgu} 
  J.~Abedi, H.~Dykaar and N.~Afshordi,
  ``Echoes from the Abyss: Tentative evidence for Planck-scale structure at black hole horizons,''
  Phys.\ Rev.\ D {\bf 96}, no.~8, 082004 (2017)
  [arXiv:1612.00266 [gr-qc]].

\bibitem{Westerweck:2017hus} 
  J.~Westerweck {\it et al.},
  ``Low significance of evidence for black hole echoes in gravitational wave data,''
  Phys.\ Rev.\ D {\bf 97}, no.~12, 124037 (2018)
  [arXiv:1712.09966 [gr-qc]].

\bibitem{Lo:2018sep} 
  R.~K.~L.~Lo, T.~G.~F.~Li and A.~J.~Weinstein,
  ``Template-based Gravitational-Wave Echoes Search Using Bayesian Model Selection,''
  Phys.\ Rev.\ D {\bf 99}, no.~8, 084052 (2019)
  [arXiv:1811.07431 [gr-qc]].

\bibitem{Nielsen:2018lkf} 
  A.~B.~Nielsen, C.~D.~Capano, O.~Birnholtz and J.~Westerweck,
  ``Parameter estimation and statistical significance of echoes following black hole signals in the first Advanced LIGO observing run,''
  Phys.\ Rev.\ D {\bf 99}, no.~10, 104012 (2019)
  [arXiv:1811.04904 [gr-qc]].

\bibitem{Uchikata:2019frs} 
  N.~Uchikata, H.~Nakano, T.~Narikawa, N.~Sago, H.~Tagoshi and T.~Tanaka,
  ``Searching for black hole echoes from the LIGO-Virgo Catalog GWTC-1,''
  Phys.\ Rev.\ D {\bf 100}, no.~6, 062006 (2019)
  [arXiv:1906.00838 [gr-qc]].

\bibitem{Tsang:2019zra} 
  K.~W.~Tsang, A.~Ghosh, A.~Samajdar, K.~Chatziioannou, S.~Mastrogiovanni, M.~Agathos and C.~Van Den Broeck,
  ``A morphology-independent search for gravitational wave echoes in data from the first and second observing runs of Advanced LIGO and Advanced Virgo,''
  arXiv:1906.11168 [gr-qc].

\end{thebibliography}
\end{document}